\documentclass[12pt]{iopart}
\usepackage{graphicx}

\newcommand{\be}{\begin{eqnarray}}
\newcommand{\ee}{\end{eqnarray}}
\newcommand{\ba}{\begin{array}}
\newcommand{\ea}{\end{array}}
\newcommand{\no}{\nonumber}
\def\Vec#1{\mbox{\boldmath $#1$}}

\begin{document}

\title[RSB, complexity, and spin representation in the GREM]
{Replica symmetry breaking, complexity and 
spin representation in the generalized random energy model}

\author{Tomoyuki Obuchi$^1$, Kazutaka Takahashi$^2$, and Koujin Takeda$^3$}

\address{
$^1$Department of Earth and Space Science, Faculty of Science, \\
Osaka University, Toyonaka 560-0043, Japan

$^2$Department of Physics, 
Tokyo Institute of Technology, Tokyo 152-8551, Japan

$^3$
Department of Computational Intelligence and Systems Science, \\
Tokyo Institute of Technology, Yokohama 226-8502, Japan
}
%\ead{custserv@iop.org}
\begin{abstract}
 We study the random energy model with a hierarchical structure 
 known as the generalized random energy model (GREM). 
 In contrast to the original analysis 
 by the microcanonical ensemble formalism, 
 we investigate the GREM by the canonical ensemble formalism
 in conjunction with the replica method. 
 In this analysis, spin-glass-order parameters are defined for 
 the respective hierarchy level,
 and all possible patterns of replica symmetry breaking (RSB)
 are taken into account. 
 As a result, we find that the higher step RSB ansatz is useful
 for describing spin-glass phases in this system.
 For investigating the nature of the higher step RSB,
 we generalize the notion of complexity developed for 
 the one-step RSB to the higher step, and demonstrate 
 how the GREM is characterized by the generalized complexity.
 In addition, we propose a novel mean-field spin-glass model
 with a hierarchical structure, 
 which is equivalent to the GREM at a certain limit.
 We also show that the same hierarchical structure
 can be implemented to other mean-field spin models than the GREM.
 Such models with hierarchy
 exhibit phase transitions of multiple steps in common.
\end{abstract}

%%75.10.Nr Spin-glass and other random models
%%64.60.De Statistical mechanics of model systems (Ising model, Potts
%%model, field-theory models, Monte Carlo techniques, etc.)
%%05.70.Fh Phase transitions: general studies
\maketitle

%%%%%%%%%%%%%%%%%%%%%%%%%%%%%%%%%%%%%%%%%%%%%%%%%%%%%%%%%%%%%%%%%%%%%%%%%%%%%%%%%%
%%%%%%%%%%%%%%%%%%%%%%%%%%%%%%%%%%%%%%%%%%%%%%%%%%%%%%%%%%%%%%%%%%%%%%%%%%%%%%%%%%
\section{Introduction}
\label{intro}

 Random spin systems exhibit rich interesting properties 
 which are absent in pure systems~\cite{MPV, Nishimori, MM}.
 Among studies of various random spin models, 
 the analysis of the Sherrington-Kirkpatrick (SK) model~\cite{SK}
 by mean-field theory 
 has revealed the essence of spin glasses, 
 where replica symmetry breaking (RSB) is significant for
 characterizing the nature of low temperature region
 as shown by Parisi's seminal works~\cite{Parisi1, Parisi2, Parisi3}.
 Recently, the RSB has attracted renewed interest 
 for some reasons.
 One of the reasons is that the solution by the Parisi ansatz
 was shown to be exact for the SK model~\cite{Guerra, Talagrand},
 which justifies the analysis by the RSB.
 Another is that the RSB was found to have a deep relation 
 to the notion of complexity~\cite{MM, BM, Monasson},
 which describes multi-valley landscape of free energy.

 In random spin systems with glassy phase, 
 it is known that free energy has a multi-valley structure, where
 each valley is sometimes called a pure state, and 
 the RSB solution is expected to describe such multi-valley structure.
 For characterizing multi-valley structure quantitatively,
 the notion of complexity was introduced as the number counting of 
 free energy minima.
 Historically, complexity was first proposed
 in terms of the Thouless-Anderson-Palmer (TAP) equation~\cite{TAP}
 by Bray and Moore in their pioneering work~\cite{BM}, 
 where they calculated the number
 of the solutions of the TAP equation.
 On the other hand, another approach for complexity
 by using the replica method was proposed by Monasson~\cite{Monasson}.
 To extract information of pure states, 
 he introduced a partition function written 
 by the copies of physical systems with 
 a weak pinning field for choosing one of pure states,
 which is in accordance with
 the replica formalism accompanied by the one-step RSB (1RSB) ansatz.
 This formulation provides not only a useful scheme to calculate complexity 
 but also a new interpretation for the 1RSB ansatz 
 from the viewpoint of complexity.
 Although this complexity is seemingly different from the one
 by number counting of the TAP solutions,
 some equivalence between them has been argued by refining 
 discussions of number counting of the TAP 
 solutions~\cite{CGG, CGPM, ACGPT, CLPR1, CLPR2, CLPR3, TTK}. 
 Currently, it is known that the complexity by the 1RSB ansatz 
 corresponds to that by the TAP framework employing 
 the Becchi-Rouet-Stora-Tyutin symmetry. 
 However, in higher step RSB systems like the SK model, 
 the correspondence between the TAP and replica complexities is 
 completely unclear. 
 This is because, for higher-step RSB systems, in the TAP context 
 the Becchi-Rouet-Stora-Tyutin symmetry leads to an incorrect solution 
 and in the replica framework 
 there is no conclusive method to calculate complexity. 
 
 The 1RSB formulation of complexity analysis
 neglects further hierarchical structure consisting of valleys,
 which is essential for describing the higher step RSB.
 Hence, it is natural 
 to generalize the notion of complexity for investigating
 hierarchical structure of valleys,
 where the higher step RSB formalism is required.
 Complexity of the higher step RSB 
 was investigated by the TAP equation formalism 
 in some preceding works~\cite{ACGPT, CLPR2, CLPR3}.
 However, they mainly focused on the family of the SK model,
 where the complexity analysis is in general difficult because
 the full-step RSB description is required in the low temperature region.

 One of the objectives of this paper is to provide a generalized
 framework of complexity for hierarchical valley structure. 
 However, before arguing generalized complexity of hierarchical valleys,
 a simple and tractable spin-glass model exhibiting
 a hierarchical valley structure
 is desired for examining application of generalized complexity 
 analysis after formulation.
 Among many models we first hit on the random energy model (REM)
 proposed by Derrida~\cite{Derrida1, Derrida2},
 which exhibits glassy nature and is exactly dealt with.
 In the original framework of the REM, spin variables do not appear
 and each energy level is drawn independently from Gaussian 
 distribution. It was also discussed that the REM is equivalent to 
 the $p$-body interacting random Ising spin model 
 at the limit $p\to\infty$, 
 which is a spin representation of the REM in some sense.
 This spin representation makes it clearer  
 how the REM is interpreted in the theory of spin glasses. 
 The replica method was applied 
 to the spin representation of the REM and the glassy phase 
 is found to be described by the 1RSB ansatz. 
 Using this correspondence, other several studies for
 the REM revealed physical meaning of the 1RSB ansatz.
 These clarified the feature of the REM 
 as the `simplest spin glass'~\cite{GM}. 
 Nevertheless, description of many other spin-glass models
 such as the SK model requires the full-step RSB ansatz, which 
 implies that the REM is not sufficient for inclusive 
 understanding of spin glasses.

 Based on such results, Derrida proposed a novel REM with
 a hierarchical structure~\cite{Derrida3, DG} known as
 the generalized random energy model (GREM).
 In the GREM, the hierarchical structure in the construction 
 of random energy levels yields
 phase transitions of multiple steps, 
 which implies the higher step RSB picture. 
 However, some problems remain unresolved.
 First, from the detailed observation
 it is found that these multiple-step phase transitions 
 are different from those of the SK model.
 Second, although the hierarchy is introduced in this model,
 it is not clear how such hierarchical structure of energy 
 is represented by means of spins, which is necessary for
 comparison with properties of other random spin models such as the SK.
 Although there is an attempt to introduce 
 the spin representation in \cite{Saakian},
 the equivalence has not been explicitly demonstrated.
 Third, this model was exactly solved by the microcanonical ensemble
 formalism but the analysis by the replica method has not been applied. 
 This leaves some ambiguities in the RSB structure of the GREM, 
 which becomes crucial for generalized formalism of complexity.

 In this paper, to give answer to the above-mentioned problems 
 we first investigate the GREM by applying the canonical ensemble
 formalism in conjunction with the replica method.
 We show that spin-glass order parameter is defined 
 for respective level of hierarchy, and 
 the 1RSB picture is realized
 for each hierarchical level.
 However, the hierarchical structure of the model 
 allows us to interpret the 1RSB solution in 
 each hierarchical level as the higher step RSB one as a whole.
 Next, after having the higher step RSB solution, 
 we propose the generalized formalism of complexity
 for hierarchical valley structure or the higher step RSB picture,
 and apply it to the GREM.
 Using the solution of the GREM, we demonstrate
 how the generalized complexity for hierarchical valley 
 is calculated and the phase transition is
 characterized in terms of the generalized complexity.
 In the end, from the solution of the GREM by
 the canonical ensemble formalism,
 we propose a spin model with hierarchical structure,
 from which the GREM can be reduced by taking a certain limit.
 In addition, from the spin representation of the GREM, we introduce
 a family of spin models with the hierarchical structure and
 show that if the model without hierarchy exhibits a phase transition 
 the corresponding hierarchical model can have
 a partially ordered state.
 We also discuss common properties of such hierarchical models.

 The organization of this paper is as follows.
 In section~\ref{energy}, 
 we introduce the GREM and analyze its thermodynamic property 
 with the canonical ensemble formalism.
 We utilize the replica method to study 
 how the RSB is realized in this model.
 In section~\ref{spin}, we propose the spin representation of the GREM,
 which is shown to be equivalent to the original GREM
 at a certain limit.
 Then, we provide the generalized formalism of 
 complexity in section~\ref{complexity} and
 also demonstrate how the generalization of complexity is
 useful for characterizing the GREM.
 In section~\ref{hierarchy}, 
 we propose a family of hierarchical spin models
 from the spin representation of the GREM, and 
 investigate common properties of such hierarchical models.
 Section~\ref{conclusion} is devoted to conclusions.

%%%%%%%%%%%%%%%%%%%%%%%%%%%%%%%%%%%%%%%%%%%%%%%%%%%%%%%%%%%%%%%%%%%%%%%%%%%%%%%%%%
%%%%%%%%%%%%%%%%%%%%%%%%%%%%%%%%%%%%%%%%%%%%%%%%%%%%%%%%%%%%%%%%%%%%%%%%%%%%%%%%%%
\section{The GREM: (1) energy representation}
\label{energy}

 We start our discussion from the original definition of the GREM.
 Then, we study the thermodynamic state of the model
 by exploiting the canonical ensemble,
 as opposed to the original analysis by the microcanonical ensemble.
 To handle the average over quenched randomness, 
 we use the replica method, which allows us to see 
 how the phase transition is characterized by the RSB solution.

%%%%%%%%%%%%%%%%%%%%%%%%%%%%%%%%%%%%%%%%%%%%%%%%%%%%%%%%%%%%%%%%%%%%%%%%%%%%%%%%%%
\subsection{Model}

 We consider a hierarchical structure of energy levels.
 The number of hierarchy levels is denoted by $K$.
 To the $\nu$th level of hierarchy ($1\le\nu\le K$)
 we assign random variables 
 $\epsilon_{\nu}(1), \epsilon_{\nu}(2), \cdots, \epsilon_{\nu}(M_\nu)$.
 The number of variables $M_\nu$ is given by 
\be
 M_\nu= (\alpha_1\cdots\alpha_\nu)^N, \label{mnu}
\ee
 where $\alpha_\nu^N$ are integer with 
 $1<\alpha_\nu^N< 2^N$ satisfying
\be
 (\alpha_1\cdots\alpha_K)^N=2^N. \label{alpha}
\ee
 We have $M_K=2^N$, which means that the number of variables 
 at the deepest hierarchy level $K$ is always equal to $2^N$.

 For the $\nu$th level of hierarchy,
 we generate random numbers with Gaussian distribution
\be
 P_\nu(\epsilon_\nu) = \frac{1}{\sqrt{\pi NJ^2 a_\nu}}
 \exp\left(-\frac{\epsilon_\nu^2}{NJ^2 a_\nu}\right).
 \label{gauss}
\ee
 Then, the variance of the random numbers is given by 
\be
 [\epsilon_\nu^2] = \frac{NJ^2}{2}a_\nu,
\ee
 where the square brackets $[\ ]$ denote the average 
 over the quenched randomness, and $a_\nu>0$ satisfies 
\be
 \sum_{\nu=1}^K a_\nu = 1. \label{suma}
\ee
 Condition (\ref{suma}) is important 
 when we see the correspondence between the GREM and the standard REM.

 From the random numbers generated as above,
 we construct $2^N$-random numbers as
\be
 E_i 
 = \sum_{\nu=1}^K\epsilon_{\nu}^{(i)}
 = \sum_{\nu=1}^K\epsilon_{\nu}(\lfloor(i-1)M_\nu/2^N \rfloor+1),
 \label{Ei}
\ee
 where $i=1,2,\cdots, 2^N$ and $\lfloor x \rfloor$ is the floor function
 which indicates the largest integer not exceeding $x$.

 The GREM is a system with energy levels (\ref{Ei}), and
 the model without hierarchy ($K=1$) is
 the standard REM~\cite{Derrida1, Derrida2}.
 This model undergoes phase transitions at some critical points
 as decreasing temperature, and possible values of critical 
 temperature are given as follows.
 A critical temperature of the whole system,
 which separates a low temperature ordered phase 
 from a high temperature disordered one, 
 is defined by 
\be
 T_{\rm c} = \frac{J}{2\sqrt{\ln 2}}.
\ee
 We also define a critical temperature of each hierarchy level by 
\be
 T_\nu=\frac{J}{2}\sqrt{\frac{a_\nu}{\ln \alpha_\nu}}, 
 \label{Tnu}
\ee
 where $\nu=1,2,\cdots, K$.
 However, they do not mean that phase transitions 
 always occur at these temperatures.
 The number of transition points depends on the values
 of the hierarchy parameters $a_\nu$ and $\alpha_\nu$~\cite{Derrida3}.
 For instance, for $K=2$, if we choose the hierarchy 
 parameters such that $T_2<T_1$
 we have two phase transitions at $T_1$ and $T_2$.
 On the other hand,
 for $T_1\le T_2$, the transition occurs only at $T=T_{\rm c}$, 
 which is the same as the REM.
 For $K\ge 3$, we may observe multiple-step transitions 
 for suitable values of the hierarchy parameters~\cite{DG}.
 In the following calculation we mainly focus on the $K=2$ case.

%%%%%%%%%%%%%%%%%%%%%%%%%%%%%%%%%%%%%%%%%%%%%%%%%%%%%%%%%%%%%%%%%%%%%%%%%%%%%%%%%%
\subsection{Replica method}

 For a given configuration, the partition function is expressed by
\be
 Z = \sum_{i=1}^{2^N} e^{-\beta E_i}
 = \sum_{i=1}^{2^N} \exp
 \left(-\beta\sum_{\nu=1}^K \epsilon_{\nu}^{(i)}\right),
\ee
 where $\beta=1/T$ is the inverse temperature.
 Following the standard method~\cite{MM},  
 we introduce replicas.
 We write the $n$th power of the partition function as 
\be
 Z^n = \sum_{i_1=1}^{2^N} \cdots \sum_{i_n=1}^{2^N}
 \exp\left(
 -\beta\sum_{\nu=1}^K\sum_{j=1}^{M_\nu} n_{\nu}(j,\{i_a\})\epsilon_{\nu}(j)
 \right),
\ee
 where 
\be
 n_{\nu}(j,\{i_a\})=\sum_{a=1}^n I_\nu(j,i_a),
\ee
 $I$ is the indicator function defined as
\be
 I_\nu(j,i_a) = \left\{\ba{ll}
 1 & \mbox{for}\ \ j=\lfloor (i_a-1)M_\nu/2^N \rfloor+1 \\
 0 & \mbox{otherwise}
\ea\right..
\ee
 Then, the average over the quenched randomness
 is performed and we find
\be
 [Z^n] = \sum_{i_1=1}^{2^N} \cdots \sum_{i_n=1}^{2^N}
 \exp\left(\frac{N\beta^2J^2}{4}\sum_{\nu=1}^K a_\nu
 \sum_{a,b=1}^{n}q_{\nu}^{ab}\right),
\ee
 where 
\be
 q_{\nu}^{ab} &=& \sum_{j=1}^{M_\nu}I_\nu(j,i_a)I_\nu(j,i_b)  \no\\
 &=& \left\{\ba{ll}
 1 & \mbox{for}\ \ 
 \lfloor (i_a-1)M_\nu/2^N \rfloor= \lfloor (i_b-1)M_\nu/2^N \rfloor \\
 0 & \mbox{otherwise}
 \ea\right..
 \label{sgparam}
\ee
 Thus, the partition function can be written in terms of 
 order parameters $q_\nu^{ab}$
 defined at each level of hierarchy.
 It follows from (\ref{sgparam}) that
 if $i_a$ and $i_b$ belong to the same group 
 at the $\nu$th level, $q_{\nu}^{ab}=1$, and zero otherwise.
 Then, we can rewrite $[Z^n]$ as 
\be
 \label{Zn}
 [Z^n]= 
 \sum_{\{q_\nu^{ab}\}} \exp\left(
 S_q(q)+\frac{N\beta^2J^2}{4}
 \sum_{\nu=1}^K a_\nu \sum_{a,b=1}^nq_{\nu}^{ab}
 \right).
\ee
 $S_q(q)$ is the entropy function 
 defined for the number of configurations giving $\{q_\nu^{ab}\}$.

%%%%%%%%%%%%%%%%%%%%%%%%%%%%%%%%%%%%%%%%%%%%%%%%%%%%%%%%%%%%%%%%%%%%%%%%%%%%%%%%%%
\subsection{Saddle point}
\label{sec:SP}

\begin{figure}[htb]
\begin{center}
\includegraphics[width=1.0\columnwidth]{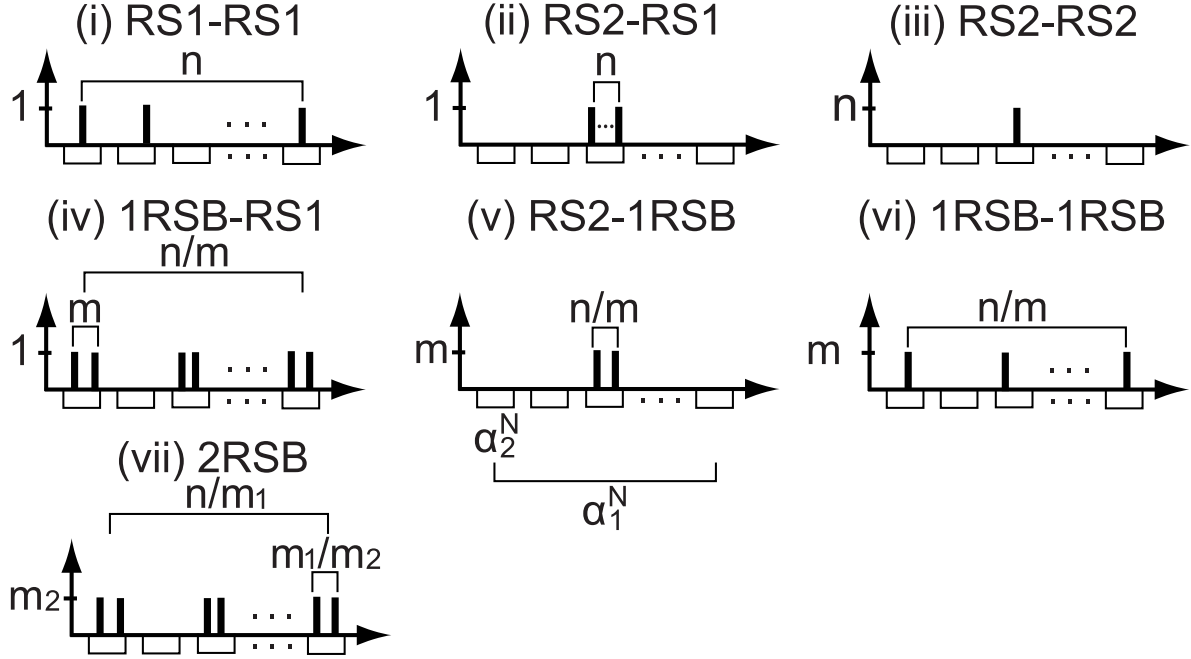}
\caption{Possible saddle point for $\{q_\nu^{ab}\}$ at $K=2$.
 The horizontal axis represents the index of configurations running 
 from 1 to $2^N$.
 All configurations are divided into $\alpha_1^N$-groups 
 including $\alpha_2^N$-configurations.
 The vertical axis represents the number of `balls'.}
\label{q}
\end{center}
\end{figure}

 Our next task is to evaluate saddle-point contribution of (\ref{Zn}).
 However, it is a difficult task to solve the saddle-point equation for 
 $\{q_\nu^{ab}\}$ generically, and here
 we obtain solutions heuristically, which can be justified 
 by the result from other analysis. 
 In the case of the REM, 
 three saddle-point solutions, called 
 (a) the replica symmetric (RS) solution of the first case (denoted by RS1),
 (b) the RS solution of the second case (RS2) 
 and (c) 1RSB solution 
 are known to exist~\cite{MM, GD, OgK1, OgK2, OgK3}.
 We extend this result to the GREM, which yields seven saddle-point
 solutions for $K=2$ as summarized below.
 These solutions can be graphically expressed by 
 how $n$-`balls' are partitioned into $2^N$-`boxes'~\cite{OgK1}.
 For $K=2$ we depict possible solutions in figure~\ref{q} and
 summarize the expressions of $S_q(q)$, $\phi(n)=(1/N)\ln[Z^n]$ 
 and thermodynamic functions as follows. 

%%%%%%%%%%%%%%%%%%%
%%%%%%%%%%%%%%%%%%%

\begin{enumerate}
\item{RS1-RS1.}

 There is no overlap between any of the configurations and 
\be
 (q_1^{ab}, q_2^{ab})=(\delta_{ab}, \delta_{ab}).
\ee
 Then, $S_q(q)$ and $\phi(n)$ are calculated as 
\be
 & & S_q(q) = \ln\left\{2^N(2^N-\alpha_2^N)\cdots 
 (2^N-(n-1)\alpha_2^N)\right\}
 \sim Nn\ln 2, \\
 & & \phi(n) = n\left(\ln 2+\frac{\beta^2J^2}{4}\right).
 \label{phin-rs1}
\ee
 This is the standard `paramagnetic' state of the REM.
 The free energy $f(T)=-\lim_{n\to 0}\phi(n)/\beta n$ 
 and the thermodynamic entropy $s(T)=-\partial f(T)/\partial T$ 
 per site $N$ can be calculated as
\be
 & & f(T) = -T\ln 2 -\frac{J^2}{4T}, \\
 & & s(T) = \ln 2 -\frac{J^2}{4T^2}. \label{spara}
\ee

\item{RS2-RS1.}

 In this case, there are overlaps at the first hierarchy and 
\be
 (q_1^{ab}, q_2^{ab})=(1, \delta_{ab}).
\ee
 We obtain 
\be
 & & S_q(q) = N\left(\ln \alpha_1+n\ln\alpha_2\right), \\
 & & \phi(n) = 
 \ln \alpha_1+\frac{\beta^2J^2}{4}n^2a_1
 +n\left(\ln\alpha_2+\frac{\beta^2J^2}{4}a_2\right).
\ee
 $\phi(n)$ is not proportional to $n$ and
 this solution is irrelevant at the limit of $n\to 0$.

\item{RS2-RS2.}

 All states have a same configuration and 
\be
 (q_1^{ab}, q_2^{ab})=(1, 1).
\ee
 We obtain 
\be
 & & S_q(q) = N\ln 2, \\
 & & \phi(n) = \ln 2+\frac{\beta^2J^2}{4}n^2.
\ee
 This is also an irrelevant solution at the limit of $n\to 0$.

\item{1RSB-RS1.}

 The replica symmetry in the first hierarchy
 is broken by introducing an integer $m$ as 
\be
 (q_1^{ab}, q_2^{ab})=(\delta_{m}(a,b), \delta_{ab}),
\ee
 where 
\be
 \delta_m(a,b) = \left\{\ba{ll} 1 & 
 \mbox{for}\ \lfloor (a-1)/m \rfloor= \lfloor (b-1)/m \rfloor \\ 
 0 &  \mbox{otherwise} \ea\right..
\ee
 Then, we have 
\be
 & & S_q(q) = Nn\left(\frac{1}{m}\ln\alpha_1+\ln\alpha_2\right), \\
 & &  \phi(n;m) = n\left(
 \frac{1}{m}\ln\alpha_1+\frac{\beta^2J^2}{4}ma_1
 +\ln\alpha_2+\frac{\beta^2J^2}{4}a_2
 \right). \label{phinm}
\ee
 $m$ is optimized so that $\phi(n)$ becomes maximum.
 We obtain a real value as $m = T/T_1$ and 
\be
 \phi(n)=n\left(\beta J\sqrt{a_1\ln\alpha_1}
  +\ln\alpha_2+\frac{\beta^2J^2}{4}a_2
 \right).
\ee
 Then, we have 
\be
 & & f(T) = -J\sqrt{a_1\ln\alpha_1}
 -T\ln \alpha_2 -\frac{J^2}{4T}a_2, \\
 & & s(T) = \ln \alpha_2 -\frac{J^2}{4T^2}a_2.
\ee

\item{RS2-1RSB.}

 The replica symmetry in the second hierarchy is broken as 
\be
 & & (q_1^{ab}, q_2^{ab})=(1, \delta_{m}(a,b)), \\
 & & S_q(q) = N\left(\ln\alpha_1+\frac{n}{m}\ln\alpha_2\right), \\
 & & 
 \phi(n;m) = 
 \ln \alpha_1+\frac{\beta^2J^2}{4}n^2a_1
 +n\left(\frac{1}{m}\ln\alpha_2+\frac{\beta^2J^2}{4}ma_2\right).
\ee
 We obtain $m = T/T_2$ and an irrelevant solution at $n\to 0$ as 
\be
 \phi(n) = \ln \alpha_1+\frac{\beta^2J^2}{4}n^2a_1 
 +n\beta J\sqrt{a_2\ln\alpha_2}.
\ee

\item{1RSB-1RSB.}
 
 The replica symmetry is broken in both the hierarchy levels as  
\be
 & & (q_1^{ab}, q_2^{ab})=(\delta_{m}(a,b), \delta_{m}(a,b)), \\
 & & S_q(q) = N\frac{n}{m}\ln 2, \\
 & & \phi(n;m) = n\left(\frac{1}{m}\ln 2
 +\frac{\beta^2J^2}{4}m \right).
\ee
 Then, $m=T/T_{\rm c}$ and the result reduces to the standard 1RSB
 `spin-glass' state of the REM as 
\be
 & & \phi(n) = n \beta J\sqrt{\ln 2}, \\
 & & f(T) = -J\sqrt{\ln 2}, \\
 & & s(T) = 0.
\ee

\item{2RSB (two-step RSB).}

 This solution is similar to the 1RSB-1RSB one but 
 the parameters $m_1$ and $m_2$ take different values as 
\be
 & & (q_1^{ab}, q_2^{ab})=(\delta_{m_1}(a,b), \delta_{m_2}(a,b)), \\
 & & S_q(q) = Nn\left(\frac{1}{m_1}\ln\alpha_1
 +\frac{1}{m_2}\ln\alpha_2\right), \\
 & & \phi(n;m_1,m_2) = 
 n\left(
 \frac{1}{m_1}\ln \alpha_1
 +\frac{\beta^2J^2}{4}m_1a_1
 +\frac{1}{m_2}\ln \alpha_2
 +\frac{\beta^2J^2}{4}m_2a_2
 \right). \no\\ \label{phi2rsb}
\ee
 We obtain $m_1 = T/T_1$, $m_2 = T/T_2$ and 
\be
 & & \phi(n) = n\left(
 \beta J\sqrt{a_1\ln\alpha_1}
 +\beta J\sqrt{a_2\ln\alpha_2}
 \right), \\
 & & f(T) = -J\sqrt{a_1\ln\alpha_1}-J\sqrt{a_2\ln\alpha_2}, \\
 & & s(T) = 0.
\ee

\end{enumerate}

 As we mentioned above, 
 the solutions (ii), (iii) and (v) are irrelevant 
 for constructing the free energy, 
 since they do not give correct behavior of $\phi(n)$ 
 at the limit $n\to 0$.
 From the other solutions, 
 we should choose a suitable solution by considering 
 the physical plausibility, 
 which leads to two situations depending on the hierarchy parameters.
 
 The first situation is the case $T_2<T_1$. 
 In this case, two phase transitions, 
 reflecting the entropy crises in each hierarchy,
 occur as temperature decreases. 
 At high temperatures, the correct solution is given by 
 RS1-RS1 solution (i) being the usual paramagnetic one. 
 As temperature decreases, at $T=T_1$ a phase transition occurs and 
 the system goes to the 1RSB-RS1 phase (iv). 
 To understand this transition, 
 we should identify the entropies of each hierarchy 
 $s_{\nu}(T)$ ($\nu=1,2$) which are given by
\be
 s_\nu(T)=\ln\alpha_\nu-\frac{J^2}{4T}a_\nu.
\ee
 The total entropy $s(T)$ is reproduced by the summation of the ones 
 in each hierarchy as $s(T)=\sum_{\nu}s_{\nu}(T)$. 
 At $T=T_1$, the entropy of the first hierarchy 
 becomes zero and the system is partially frozen to 
 its ground state in the first hierarchy, 
 which can be clearly seen in the solution (iv).
 Similarly, at $T=T_2<T_1$, the entropy of the second hierarchy 
 also becomes zero, which leads to a phase transition from (iv) 
 to the 2RSB solution (vii). 
 The above RSB transitions are hence interpreted as 
 the entropy crises in different hierarchies, 
 which sequentially occurs from the upper macroscopic to 
 lower microscopic levels of hierarchy. 
 
 The other case is for $T_1 \leq T_2$. 
 In this case, the correct solution is given by 
 (i) for $T>T_{\rm c}$  and by (vi) for $T\leq T_{\rm c}(< T_2)$. 
 This can be easily found since a possible branch at 
 low temperatures, which should continuously connect to 
 the high temperature solution (i), is only (iv) in this case. 
 This solution indicates that the hierarchical structure becomes 
 irrelevant and the standard REM result is reproduced.
 
 The solution for the case $T_1 \leq T_2$ might give a question 
 to the hierarchical entropy $s_{\nu}$ 
 since the second-hierarchy entropy $s_2(T)$ becomes 
 negative in the region $T_{\rm c}<T<T_2$. 
 This implies that $s_{\nu}(T)$ 
 has its physical significance only in the case that the entropies 
 in upper levels of hierarchy $s_\mu(T)$ with $\mu<\nu$ vanish. 
 This point requires further discussions and 
 is revisited in section~\ref{complexity}.

\begin{center}
\begin{figure}[htb]
\begin{minipage}[h]{0.5\textwidth}
\begin{center}
\includegraphics[width=0.9\columnwidth]{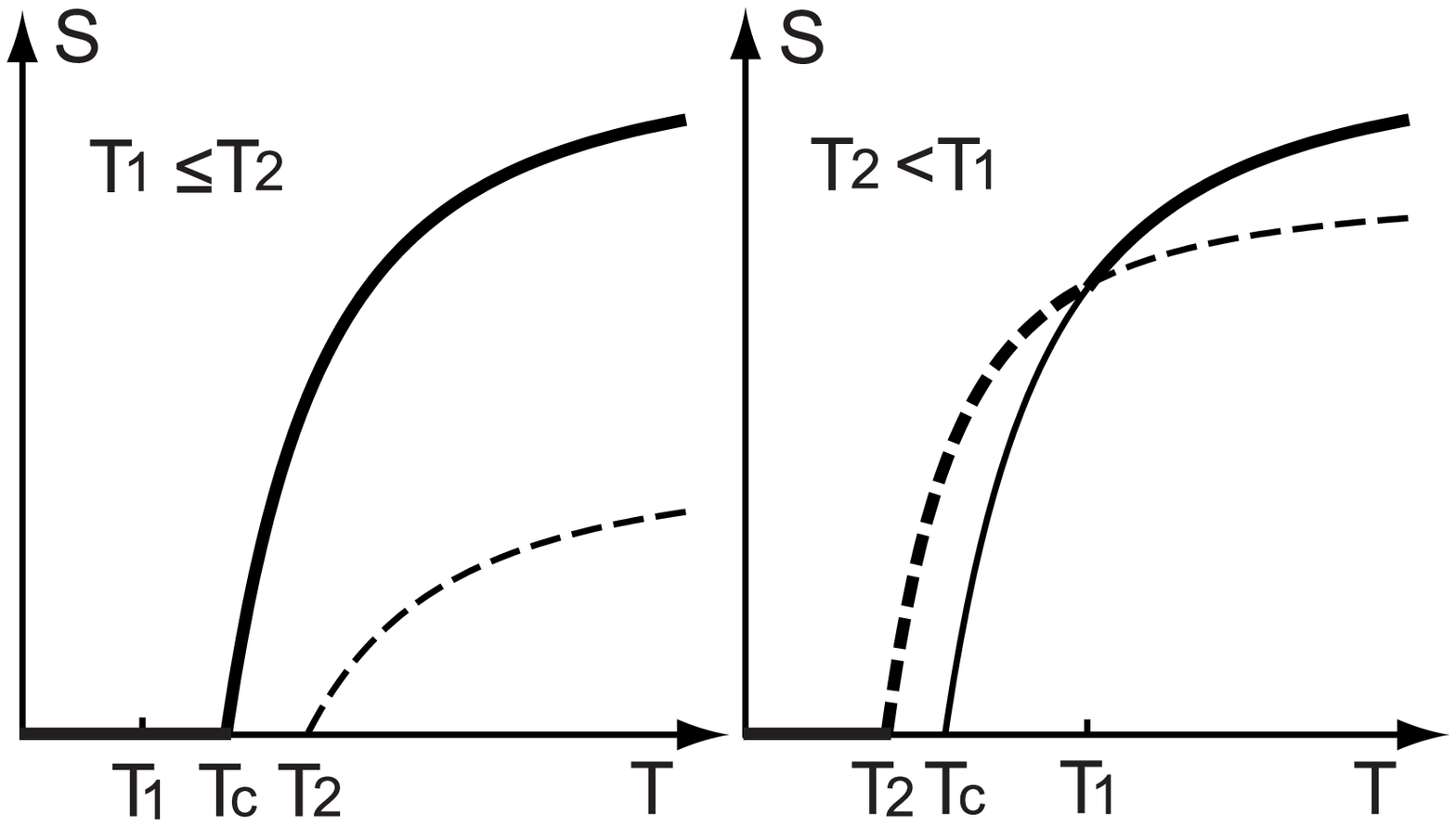}
\caption{The entropy $s(T)$ of the GREM at $K=2$. 
 The dashed line represents the entropy of the second hierarchy 
 and the solid line the total entropy.
 The actual entropy takes the bold parts.
}
\label{st}
\end{center}
\end{minipage}
\begin{minipage}[h]{0.5\textwidth}
\begin{center}
\includegraphics[width=0.9\columnwidth]{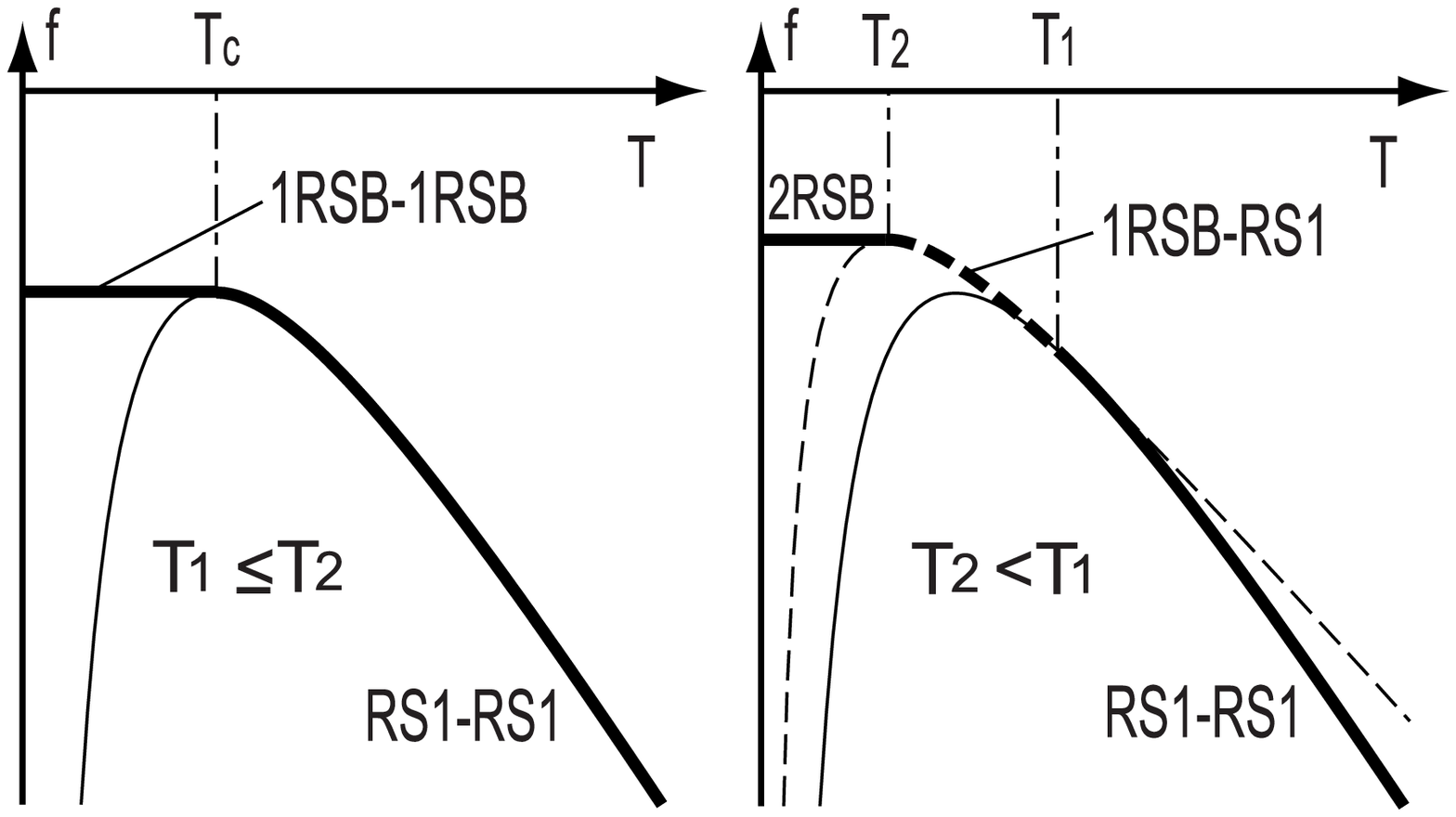}
\caption{The free energy $f(T)$ of the GREM at $K=2$.
 From among several curves representing possible phases, 
 the actual free energy takes the bold parts.}
\label{ft}
\end{center}
\end{minipage}
\end{figure}
\end{center}

 Now, we summarize the behavior of thermodynamic functions as follows.
 The schematic behaviors of the entropy and the free energy
 are also depicted in figures~\ref{st} and \ref{ft}, respectively.

%%%%%%%%%%%%%%
%%%%%%%%%%%%

\begin{itemize}
\item{$T_2<T_1$}
\be
 & & s(T)=\left\{\ba{ll} \ln2-\frac{J^2}{4T^2} & T_1\le T \\
 \ln\alpha_2-\frac{J^2}{4T^2}a_2 & T_2\le T\le T_1 \\ 
 0 & T\le T_2 \ea\right., \\
 & & f(T)=\left\{\ba{ll} -T\ln 2 -\frac{J^2}{4T} & T_1\le T \\
 -J\sqrt{a_1\ln\alpha_1}-T\ln\alpha_2-\frac{J^2}{4T}a_2 & 
 T_2\le T\le T_1 \\
 -J\left(\sqrt{a_1\ln\alpha_1}+\sqrt{a_1\ln\alpha_1}\right) 
 & T\le T_2 \ea\right..
 \label{f-T1>T2}
\ee

\item{$T_1\le T_2$}
\be
 & & s(T)=\left\{\ba{ll} \ln2-\frac{J^2}{4T^2} & T_{\rm c}\le T \\ 
 0 & T\le T_{\rm c} \ea\right., \\
 & & f(T)=\left\{\ba{ll} -T\ln 2 -\frac{J^2}{4T} & T_{\rm c}\le T \\ 
 -J\sqrt{\ln 2} & T\le T_{\rm c} \ea\right.. 
 \label{f-T1<T2}
\ee
\end{itemize}

 This result coincides with the original one 
 by the microcanonical ensemble formalism~\cite{Derrida3}.
 As we see, the phase at lower temperature 
 is described by the 2RSB for a suitable choice
 of the hierarchy parameters.
 The generalization of the discussion to $K \ge 3$ is straightforward, 
 where the higher step RSB is realized
 at lower temperature region.

%%%%%%%%%%%%%%%%%%%%%%%%%%%%%%%%%%%%%%%%%%%%%%%%%%%%%%%%%%%%%%%%%%%%%%%%%%%%%%%%%%
%%%%%%%%%%%%%%%%%%%%%%%%%%%%%%%%%%%%%%%%%%%%%%%%%%%%%%%%%%%%%%%%%%%%%%%%%%%%%%%%%%
\section{The GREM: (2) spin representation}
\label{spin}

 Our main motivation for introducing the GREM is to understand 
 spin-glass phase transitions characterized by the higher step RSB.
 However, the GREM in the previous section is defined by 
 randomly distributed energy, and the relation of the GREM
 with other random models expressed by spins is not clear.
 On the other hand, 
 the standard REM is known as the limit $p\to\infty$ of 
 the $p$-body interacting spin-glass model.
 This observation is very useful to find a connection  
 between the REM with the 1RSB and 
 the SK model with the full-step RSB.
 Therefore, it is natural to ask whether one can construct 
 a spin-glass model which includes the GREM as a limit.
 In this section, we propose a hierarchical $p$-body interacting 
 spin-glass model and
 demonstrate that this yields the same thermodynamic behavior 
 as the GREM at the limit $p\to\infty$.

%%%%%%%%%%%%%%%%%%%%%%%%%%%%%%%%%%%%%%%%%%%%%%%%%%%%%%%%%%%%%%%%%%%%%%%%%%%%%%%%%%
\subsection{Hierarchical $p$-body interacting spin-glass model}

 We define the Hamiltonian
\be
 H = -\sum_{\nu=1}^K\sum_{(i_1\cdots i_p)}^{N_\nu}
 J_{i_1\cdots i_p}^{(\nu)}\sigma_{i_1}\cdots\sigma_{i_p},
 \label{grem-s}
\ee
 where $\sigma_i$ is the Ising spin on site $i$.
 In the $\nu$th hierarchy,
 the sum is taken over possible combinations of $N_\nu$-spins,
 where $N_\nu$ is given by $2^{N_\nu}=M_\nu$ and
 $M_\nu$ is in (\ref{mnu}). 
 Due to the properties (\ref{mnu}) and (\ref{alpha}), 
 $N_1<N_2<\cdots<N_K=N$.
 The random interaction $J_{i_1\cdots i_p}^{(\nu)}$ 
 distributes in Gaussian with the variance 
\be
 [(J_{i_1\cdots i_p}^{(\nu)})^2]
 = \frac{NJ^2}{2}a_\nu \frac{p!}{N_\nu^p},
\ee
 where $a_\nu$ satisfies (\ref{suma}).

%%%%%%%%%%%%%%%%%%%%%%%%%%%%%%%%%%%%%%%%%%%%%%%%%%%%%%%%%%%%%%%%%%%%%%%%%%%%%%%%%%
\subsection{Energy-level distributions}

 Before calculating the thermodynamic functions,
 we examine the energy-level distribution functions.
 For a given configuration $\{\sigma_i\}$,
 the single-level distribution is calculated as 
\be
 \left[\delta \left(E-H(\{\sigma_i\})\right)\right] &=&
 \int\frac{dt}{2\pi}e^{itE}\left[
 \exp\left(it\sum_{\nu=1}^K\sum_{(i_1\cdots i_p)}^{N_\nu}
 J_{i_1\cdots i_p}^{(\nu)}\sigma_{i_1}\cdots\sigma_{i_p}
 \right)\right] \no\\
% &=& \int\frac{dt}{2\pi}e^{itE}
% \exp\left(-\frac{Nt^2J^2}{4}\right) \no\\
 &=& \frac{1}{\sqrt{\pi NJ^2}} 
 \exp\left(-\frac{E^2}{NJ^2}\right).
\ee
 Thus, the single energy-level distribution is equivalent to 
 the REM/GREM.
 In the same way, the two-point correlation 
 for configurations $\{\sigma_i^1\}$ and $\{\sigma_i^2\}$ is 
 calculated as 
\be
 & & \left[ \delta \left(E-H(\{\sigma_i^1\})\right) 
 \delta\left(E'-H(\{\sigma_i^2\})\right)\right]  \no\\
 &=& \frac{1}{\pi NJ^2\sqrt{1-v^2}}
 \exp\left\{
 -\frac{(E+E')^2}{2NJ^2(1+v)}
 -\frac{(E-E')^2}{2NJ^2(1-v)}
 \right\},
\ee
 where 
\be
 & & v = \sum_{\nu=1}^K a_\nu (q_\nu^{12})^p, \\
 & & (q_\nu^{12})^p 
 = \frac{p!}{N_\nu^p}\sum_{(i_1\cdots i_p)}
 \sigma_{i_1}^1\sigma_{i_1}^2\cdots\sigma_{i_p}^1\sigma_{i_p}^2
 \sim \left(\frac{1}{N_\nu}\sum_{i=1}^{N_\nu}\sigma_{i}^1\sigma_{i}^2\right)^p.
\ee
 This form is the same as that of the GREM~\cite{Derrida3}. 
 The quantity $v$ depends on the spin configurations and 
 it goes to the result of the GREM at the limit $p\to\infty$.

%%%%%%%%%%%%%%%%%%%%%%%%%%%%%%%%%%%%%%%%%%%%%%%%%%%%%%%%%%%%%%%%%%%%%%%%%%%%%%%%%%
\subsection{Replica method}

 We calculate the ensemble average of 
 the $n$th power of the partition function.
 Introducing the replica indices $a=1,2,\cdots, n$, we obtain 
\be
 [Z^n] &=& \left[
 \Tr \exp\left\{
 \sum_{\nu=1}^K\sum_{(i_1\cdots i_p)}^{N_\nu}
 \beta J_{i_1\cdots i_p}^{(\nu)}\sum_{a=1}^n\sigma_{i_1}^a\cdots \sigma_{i_p}^a
 \right\}\right] \no\\
 &=& 
 \Tr \exp\left\{
 \frac{N\beta^2J^2}{2}
 \sum_{\nu=1}^Ka_\nu\sum_{a> b}^{n}
 \left(\frac{1}{N_\nu}\sum_{i=1}^{N_\nu}\sigma_{i}^a\sigma_{i}^b\right)^p
 +\frac{Nn\beta^2J^2}{4}
 \right\}.
\ee
 The order parameters 
 $q_\nu^{ab}\sim \sum_{i=1}^{N_\nu}\sigma_{i}^a\sigma_{i}^b/N_\nu$
 and auxiliary variables $\tilde{q}_\nu^{ab}$ are 
 introduced following the standard prescription as 
\be
 [Z^n]
 &=& \int dq_\nu^{ab}\Tr 
 \prod_{\nu=1}^K\prod_{a>b}^n\delta\left(
 q_\nu^{ab}-\frac{1}{N_\nu}\sum_{i=1}^{N_\nu}\sigma_{i}^a\sigma_{i}^b
 \right)
 \no\\
 & & \times\exp\left\{
 \frac{N\beta^2J^2}{2}\sum_{\nu=1}^K 
 a_\nu\sum_{a>b}^n
 \left(q_\nu^{ab}\right)^p
 +\frac{Nn\beta^2J^2}{4}
 \right\} \no\\
 &=& 
 \int dq_\nu^{ab}d\tilde{q}_\nu^{ab}\Tr 
 \exp\left\{
 -N\beta^2J^2\sum_{\nu=1}^K a_\nu \sum_{a>b}^n\tilde{q}_\nu^{ab}\left(
 q_\nu^{ab}-\frac{1}{N_\nu}\sum_{i=1}^{N_\nu}\sigma_{i}^a\sigma_{i}^b
 \right)\right.
 \no\\
 & &
 \left.+\frac{N\beta^2J^2}{2}\sum_{\nu=1}^K
 a_\nu\sum_{a> b}^n
 \left(q_\nu^{ab}\right)^p
 +\frac{Nn\beta^2J^2}{4}
 \right\}.
\ee
 The integrations are evaluated by the saddle-point method.
 Using the saddle point to be obtained, 
 we can write 
\be
 [Z^n] &=&
 \exp\left\{
 -\frac{N\beta^2J^2}{2}(p-1)\sum_{\nu=1}^K a_\nu\sum_{a>b}^n(q_\nu^{ab})^p
 +\frac{Nn\beta^2J^2}{4}\right. 
 \no\\
 &&
 \left.
 +\ln\Tr \exp\left(N\beta^2J^2\sum_{\nu=1}^K\frac{a_\nu}{N_\nu}\sum_{a>b}^n
 \tilde{q}_\nu^{ab}\sum_{i=1}^{N_\nu} \sigma_i^a\sigma_i^b
 \right)\right\}, 
\ee
 where 
\be
 \tilde{q}_\nu^{ab}=\frac{p}{2}(q_\nu^{ab})^{p-1}.
\ee

 In the following calculation, we set $K=2$ for simplicity. 
 Then, the partition function reads 
\be
 [Z^n] &=& 
 \exp\left\{
 -\frac{N\beta^2J^2}{2}(p-1)\sum_{\nu=1}^2 a_\nu\sum_{a>b}^n(q_\nu^{ab})^p
 +\frac{Nn\beta^2J^2}{4}\right. 
 \no\\
 & &
 +N_1\ln\Tr \exp\left(N\beta^2J^2\sum_{\nu=1}^2\frac{a_\nu}{N_\nu}
 \sum_{a>b}^n
 \tilde{q}_\nu^{ab} \sigma^a\sigma^b\right)
 \no\\
 & & \left.
 +(N-N_1)\ln\Tr \exp\left(N\beta^2J^2\frac{a_2}{N}\sum_{a>b}^n
 \tilde{q}_2^{ab} \sigma^a\sigma^b\right)
 \right\}, \label{zn}
\ee
 and the saddle-point equation can be derived as
\be
 & & q_1^{ab} = \frac{\Tr \sigma^a\sigma^b
 \exp\left(
 N\beta^2J^2\sum_{\nu=1}^2\frac{a_\nu}{N_\nu}\sum_{a>b}^n
 \tilde{q}_\nu^{ab} \sigma^a\sigma^b
 \right)}{\Tr\exp\left(
 N\beta^2J^2\sum_{\nu=1}^2\frac{a_\nu}{N_\nu}\sum_{a>b}^n
 \tilde{q}_\nu^{ab}\sigma^a\sigma^b
 \right)}, \\
 & & q_2^{ab} = \frac{N_1}{N}q_1^{ab}
 +\frac{N-N_1}{N}
 \frac{\Tr \sigma^a\sigma^b
 \exp\left(
 \beta^2J^2a_2\sum_{a>b}^n
 \tilde{q}_2^{ab} \sigma^a\sigma^b
 \right)}{\Tr\exp\left(
 \beta^2J^2a_2\sum_{a>b}^n
 \tilde{q}_2^{ab} \sigma^a\sigma^b
 \right)}.
\ee
 The saddle-point equations can be easily solved at $p=\infty$ 
 since the possible solutions of $\tilde{q}^{ab}_{1,2}$ are restricted 
 to 0 or $\infty$.
 We can repeat the similar discussion as that of the previous section.
 For example, the RS1-RS1 solution 
 $(q_1^{ab}, q_2^{ab})=(\delta_{ab}, \delta_{ab})$
 gives the same result (\ref{phin-rs1}).
 In \ref{replica}, 
 we study the 1RSB-RS1 and the 2RSB solutions 
 as nontrivial cases. 
 All cases give the same result as that of the previous section,  
 which shows that the present hierarchical spin model at $p\to\infty$ 
 is equivalent to the GREM.

 Here, we must mention a similarity between our model and 
 the diluted generalized random energy model 
 proposed by Saakian in \cite{Saakian}.
 Although the hierarchical structure is the same for both the models, 
 the form of the Hamiltonian is slightly different.
 In addition, he has not shown explicitly that 
 his model is equivalent to the GREM.
 In principle, it might be possible to define 
 several hierarchical spin models
 which reduce to the GREM at a certain limit.
 We consider that our model is the simplest one among such models. 

 We also see that our model is very different from 
 other spin models with the higher step RSB such as the SK model. 
 Each spin is not treated in an equivalent manner, 
 which is clearly the origin of the hierarchical ordering.
 As a random spin model with a hierarchy, 
 similar models on a hierarchical lattice
 are proposed in \cite{FJP, CDFMP}.
 These models are shown to be useful to study RSB solutions.
 The advantage of our model is that it can be treated 
 by the mean-field theory and that the analytical result is available.
 We further discuss this point
 by considering similar hierarchical models in section~\ref{hierarchy}.

%%%%%%%%%%%%%%%%%%%%%%%%%%%%%%%%%%%%%%%%%%%%%%%%%%%%%%%%%%%%%%%%%%%%%%%%%%%%%%%%%%
%%%%%%%%%%%%%%%%%%%%%%%%%%%%%%%%%%%%%%%%%%%%%%%%%%%%%%%%%%%%%%%%%%%%%%%%%%%%%%%%%%
\section{Generalization of complexity}
\label{complexity}

 As we see, it is shown that lower temperature phases in the GREM
 are described by the higher step RSB.
 Here, to investigate such systems in detail, 
 we generalize the assessment scheme of complexity for
 higher step RSB systems.
 To this end, we start from a brief review of 
 how the concept of complexity is introduced.

 The concept of complexity is closely related to 
 the multi-valley structure of the phase space.
 In a number of systems with quenched randomness such as spin glasses,  
 the phase space is considered to be divided 
 into exponentially many disjoint sets 
 in the thermodynamic limit~\cite{MRS, ObK}.
 Each component of the disjoint sets is sometimes called pure state and 
 is schematically described by a valley in the phase space. 
 Each pure state labeled by $\gamma$
 has its own free energy value $f_{\gamma}$.
 The number of pure states having the free energy value $f$, 
 $\mathcal{N}(f)$, is scaled as 
\be
 \mathcal{N}(f)\sim e^{N\Sigma(f)},
\ee 
 where the exponent $\Sigma(f)\sim O(1)$ is called 
 complexity. Note that an inequality $\Sigma(f)\geq 0$ holds since 
 complexity is the logarithm of the number of states.
 The partition function of the whole system is written 
 by the summation of the weight of each pure state $\gamma$ as 
 $Z=\sum_{\gamma} Z_\gamma=\sum_{\gamma} e^{-N\beta f_{\gamma}}$.
 This total partition function can also be written 
 by using complexity as
\be
 Z \sim \int df e^{N\left(-\beta f+\Sigma(f)\right)}.
\ee
 The saddle-point method yields the equilibrium free energy 
 $f_{\rm eq}=-(1/N\beta)\ln Z$ as
\be 
 -\beta f_{\rm eq}=\max_{f_- \leq f\leq f_+}
 \left\{-\beta f+\Sigma(f)\right\}.  \label{f_eq}
\ee
 The inequality $\Sigma(f)\geq 0$ leads to 
 the upper and lower bounds of $f$, $f_+$ and $f_-$, respectively. 
 Equation (\ref{f_eq}) means that complexity of the system 
 is necessary to obtain the equilibrium free energy. 
 This is quite general; however, the actual evaluation scheme of complexity 
 depends on the number of RSB steps.

%%%%%%%%%%%%%%%%%%%%%%%%%%%%%%%%%%%%%%%%%%%%%%%%%%%%%%%%%%%%%%%%%%%%%%%%%%%%%%%%%%
\subsection{The 1RSB case}

\begin{center}
\begin{figure}[htb]
\begin{minipage}[h]{0.5\textwidth}
\begin{center}
\includegraphics[width=0.9\columnwidth]{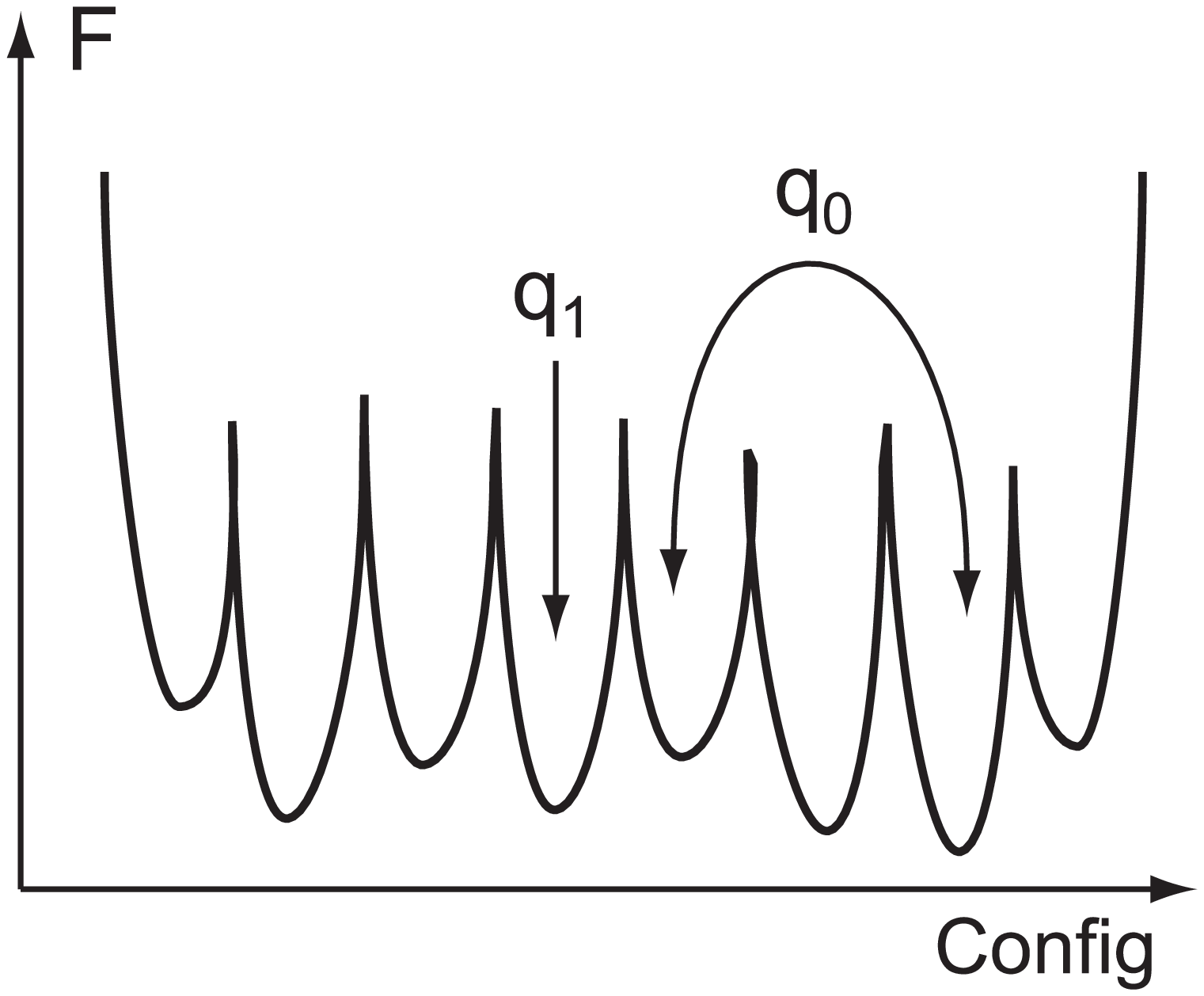}
\caption{Schematic picture of the phase space in the 1RSB phase. 
 The vertical axis represents the free energy and the horizontal 
 one schematically describes the configuration space 
 of dynamical variables. 
 Each valley corresponds to a pure state.
 There is no meta-structure constituted by pure states.}
\label{1RSB}
\end{center}
\end{minipage}
\begin{minipage}[h]{0.5\textwidth}
\begin{center}
\includegraphics[width=0.9\columnwidth]{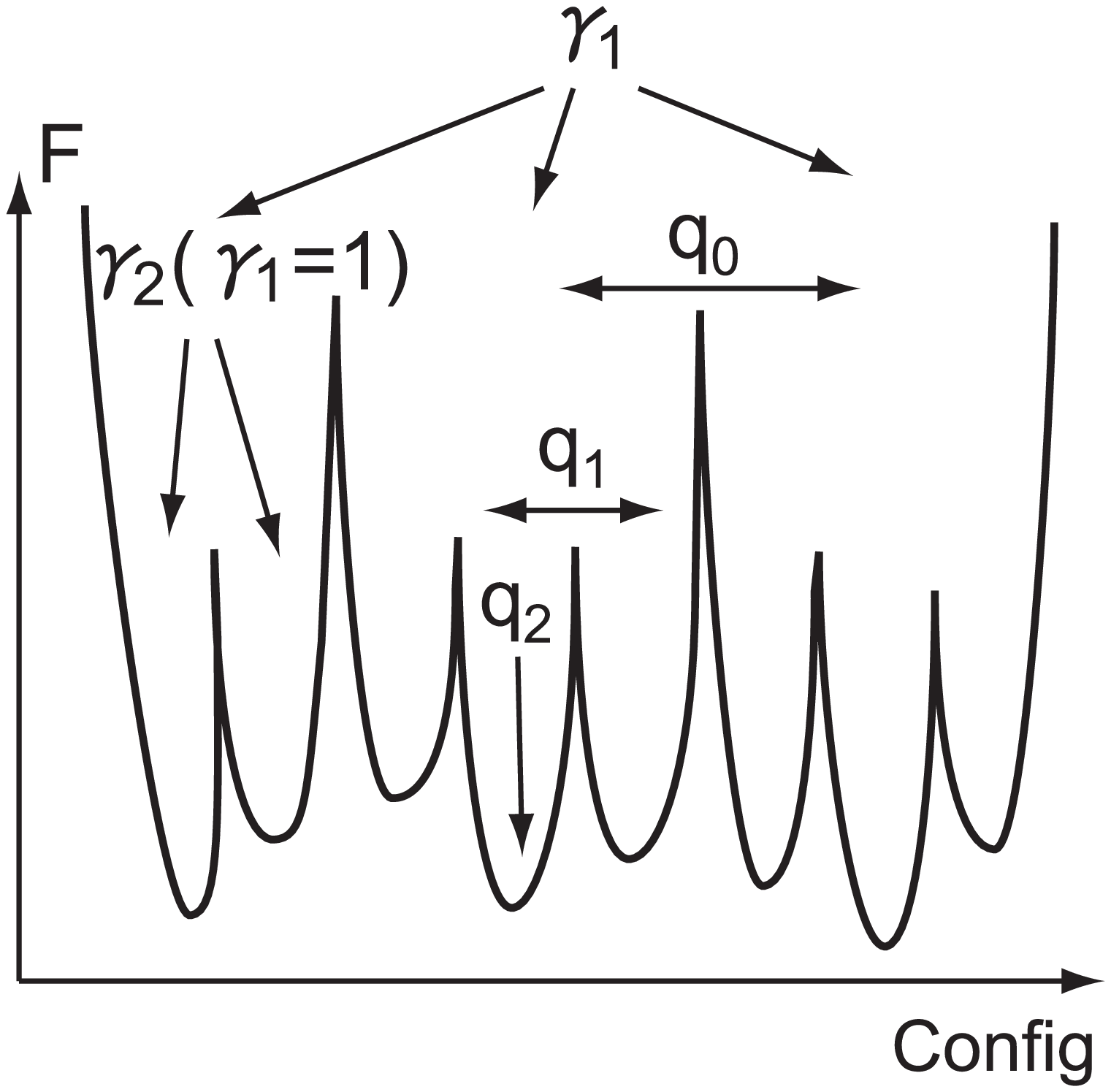}
\caption{Schematic picture of the phase space in the 2RSB phase. 
 Some valleys constitute a large `valley' in a more macroscopic level.
 Three overlap parameters
 are necessary to characterize this phase-space structure.}
\label{2RSB}
\end{center}
\end{minipage}
\end{figure}
\end{center}

 To evaluate complexity, we need some information about 
 the phase-space structure of the system. 
 In the 1RSB case, it is considered that 
 the phase space has many valleys corresponding to pure states.
 However, its structure is not complicated because
 all the pure states are statistically equivalent and 
 any meta-structure is not present in the phase space 
 (see figure~\ref{1RSB}). 
 In such a situation, it is natural to characterize 
 the phase-space structure by two typical overlaps: 
 the one in single pure state $q_1$ and 
 the other one between pure states $q_0$. 
 These are nothing but the 1RSB spin-glass-order parameters.
 This implies that the 1RSB free energy $f_{\rm 1RSB}(m)$ parameterized 
 by the breaking parameter $m$ has some information about 
 the phase space. 
 Actually, according to Monasson's argument~\cite{Monasson},  
 the 1RSB free energy is proportional to a generating function of 
 complexity $g(x)$ as $f_{\rm 1RSB}(m=x)=-g(x)/\beta x$. 
 The definition of $g(x)$ is given by 
\be
 g(x)=\frac{1}{N} \ln \left(\sum_{\gamma}(Z_{\gamma})^x\right) =
 \max_{f_{-}\leq f\leq f_{+}}
 \left\{ -\beta x f+ \Sigma(f)\right\}. \label{gx}
\ee
 Using this equation, we can calculate complexity 
 by Legendre transformation
\be
 \Sigma(f)=\min_{x}\left\{g(x)+\beta x f\right\}. \label{sigma-1RSB}
\ee
 If the complexity is convex and analytic, 
 we can express the pure-state free energy and complexity 
 in parameterized forms as 
\be 
 & & -\beta f(x) = \frac{\partial g}{\partial x}, \\
 & & \Sigma(x) = g(x)-x \frac{\partial g}{\partial x}.
\ee
 Once we obtain complexity $\Sigma(f)$ from $g(x)$ by (\ref{sigma-1RSB}), 
 the equilibrium free energy 
 can be calculated by (\ref{f_eq}). 
 These procedures constitute the scheme of how to assess complexity 
 and equilibrium free energy in the 1RSB level. 

 Before proceeding to the 2RSB case, we mention the equivalence between 
 the generating function $g(x)$ and the 1RSB free energy $f_{\rm 1RSB}(m)$.
 We can expect the self-averaging property of the generating function 
 $g(x)$, which is defined by (\ref{gx}) and intrinsically 
 depends on the quenched randomness. 
 Hence, the typical generating function $g(x)$ 
 can be replaced by the averaged one. This consideration, in conjunction 
 with the replica method, yields
\be
 g(x)= \frac{1}{N}\left[ \ln 
 \left(\sum_\gamma (Z_\gamma)^x\right) \right]
 =\lim_{y\to 0}\frac{\partial}{\partial y}
 \frac{1}{N}\ln
 \left[\left(\sum_\gamma (Z_\gamma)^x\right)^y\right].
 \label{g-replica}
\ee
 Although exact evaluation of the right-hand side of 
 (\ref{g-replica}) is difficult, 
 we can derive the following expression for integer $x$ and $y$:
\be
 \left[\left(\sum_\gamma (Z_\gamma)^x\right)^y\right]
 &=& \left[\left\{
 \sum_\gamma\left(\sum_{\{\sigma\}}
 e^{-\beta H(\{\sigma\})}\theta_\gamma (\{\sigma\})\right)^x
 \right\}^y\right] \no\\
 &=& \left[
 \prod_{\mu=1}^y\sum_{\gamma_\mu}
 \prod_{\nu=1}^x\sum_{\{\sigma_\nu^\mu\}}
 \exp\left(-\beta\sum_{\mu=1}^y\sum_{\nu=1}^x H(\{\sigma_\nu^\mu\})\right)
 \Theta(x,y)\right], \no\\ \label{phi}
\ee
 where $\{\sigma\}$ denotes the dynamical variables (spins) and 
 $\theta_{\gamma}(\{\sigma\})$ is the indicator function such that 
 $\theta_\gamma(\{\sigma\})=1$ if $\{\sigma\}\in\gamma$, and 0 otherwise. 
 We also define  
\be
 \Theta(x,y)= \prod_{\mu=1}^y\prod_{\nu=1}^x 
 \theta_{\gamma_\mu}(\{\sigma_\nu^\mu\}).
\ee
 For the evaluation of (\ref{phi}), 
 the following observations are important. 
\begin{itemize}
\item 
 The summation is taken over all possible configurations 
 of $xy$ replica spins. 
\item 
 However, the factor $\Theta(x,y)$
 allows only contributions from configurations in which $xy$ replicas 
 are equally assigned to $y$ pure states with the weight $x$. 
\end{itemize}

 These points describe nothing more than the physical picture
 of the 1RSB phase, in which case 
 we evaluate $[Z^n]$ with substitution of $n=xy$ and $m=x$.
 Hence, we obtain
\be
 & & g(x)
 = \lim_{y\to 0}\frac{\partial}{\partial y} 
 \phi_{\rm 1RSB}(n=yx,m=x)=-\beta x f_{\rm 1RSB}(x), 
 \label{phi1rsb-g}
\ee
 where $\phi_{\rm 1RSB}(n,m)$ is $(1/N) \ln[Z^n]$ 
 assessed under the 1RSB ansatz with the breaking parameter $m$. 

%%%%%%%%%%%%%%
%%%%%%%%%%%%

%%%%%%%%%%%%%%%%%%%%%%%%%%%%%%%%%%%%%%%%%%%%%%%%%%%%%%%%%%%%%%%%%%%%%%%%%%%%%%%%%%
\subsection{The 2RSB case}

 Next we consider the 2RSB case. 
 From the physical picture supposed in the 2RSB, 
 the phase space structure can be schematically depicted 
 as figure~\ref{2RSB}.
 There are two macroscopically distinct levels of hierarchies. 
 We here call them hierarchies 1 and 2~\footnote{These hierarchies 
 directly correspond to those in the $K=2$ GREM as shown later.}.
 Hierarchy 2 is constituted by pure states labeled by $\gamma_2$. 
 On the other hand, the hierarchy 1 is the one in a coarse-grained level. 
 Each coarse-grained state in this level,
 labeled by $\gamma_1$, is constituted by a group of pure states.  
 The phase space is characterized by three overlaps: 
 $q_2$ in a single pure state, 
 $q_1$ between two pure states in the same state
 of hierarchy 1 and $q_0$ between different states
 in hierarchy 1.

 Let us accept and employ the above description 
 to obtain complexity in the 2RSB level. 
 This description enables us to express the partition function as 
\be
 Z = \sum_{\gamma_1} \left(\sum_{\gamma_2(\gamma_1)} Z_{\gamma_2(\gamma_{1})}
 \right),
\ee
 where $\gamma_{1}$ specifies coarse-grained states in hierarchy 1
 and $\gamma_{2}(\gamma_{1})$ represents
 a pure state in a large valley $\gamma_1$.
 Similarly to the 1RSB case, we can introduce a generating function 
 $g_1(x_1,x_2)$ as
\be
 g_1(x_1,x_2)=\frac{1}{N}\ln \left\{
 \sum_{\gamma_1} \left(\sum_{\gamma_2(\gamma_1)} 
 (Z_{\gamma_2(\gamma_{1})})^{x_2}
 \right)^{x_1}\right\}. \label{g_1}
\ee
 Actually, this generating function $g_1(x_1,x_2)$ corresponds to 
 the 2RSB free energy with the breaking parameters 
 $m_1$ and $m_2$ as
\be
 f_{\rm 2RSB}(m_1=x_1 x_2,m_2=x_2)
 = -\frac{1}{\beta x_1x_2}g_1(x_1,x_2).
 \label{f_2RSB-g}
\ee
 This correspondence can be understood by 
 following a discussion
 being similar to that in the previous 1RSB case. 
 
 The next task is to construct a procedure to obtain complexity from 
 $g_1(x_1,x_2)$. 
 The following expression of $g_1(x_1,x_2)$ is useful
 for this purpose:
\be
 e^{Ng_1(x_1,x_2) } 
 &=& \sum_{\gamma_1}\left(\sum_{\gamma_2(\gamma_1)}
 (Z_{\gamma_2(\gamma_1)})^{x_2}\right)^{x_1} \no\\
 &=& \sum_{\gamma_1}\left(
 \int df e^{N(-\beta x_2 f+\widehat{\Sigma}_2(f|\gamma_1)) }
 \right)^{x_1}   \no\\
 &=& \int df_1 e^{N(x_1 f_1+\Sigma_{1}(f_1|x_2))},
 \label{g2RSB}
\ee
 where we introduce two entropy-like quantities 
 $\Sigma_{1}(f_1|x_2)$ and $\widehat{\Sigma}_2(f|\gamma_1)$, 
 and another generating function
 $f_1(x_2|\gamma_1)$ defined in a large valley $\gamma_1$ as 
\be 
 f_1(x_2|\gamma_1)
 =\frac{1}{N}\ln \left(
 \sum_{\gamma_2(\gamma_1)} (Z_{\gamma_2(\gamma_1)})^{x_2}\right)
 =\max_{f_{-}\leq f\leq f_{+}}
 \left\{ -\beta x_2 f +\widehat{\Sigma}_2(f|\gamma_1) \right\}.
\ee
 The physical meanings of these quantities are as follows. 
 The quantity $\widehat{\Sigma}_2(f|\gamma_1)$
 is the complexity in a large valley 
 $\gamma_1$, i.e. it characterizes the number of pure states 
 having free energy value $f$ in a valley $\gamma_1$. 
 The generating function $f_1(x_2|\gamma_1)$ plays a role of 
 a coarse-grained free energy in a valley $\gamma_1$. 
 Then, in the hierarchy 1, the number of valleys having 
 `free energy' value $f_1(x_2|\gamma_1)=f_1$, 
 $\mathcal{N}_1(f_1|x_2)$, is also important. 
 This quantity $\mathcal{N}_{1}$ intrinsically depends on 
 $x_2$ and is characterized by the exponent $\Sigma_{1}(f_1|x_2)$ 
 as $\mathcal{N}_{1}(f_1|x_2)=\exp\{N \Sigma_{1}(f_1|x_2)\}$.
 In this sense, $\Sigma_{1}(f_1|x_2)$ is  
 the `complexity of hierarchy 1'. 
 According to (\ref{g2RSB}), we can see that 
 $g_1(x_1,x_2)$ and $\Sigma_1(f_1|x_2)$ are related with each other 
 by the Legendre transformation 
\be 
 g_1 (x_1,x_2) = \max_{f_{1-}\leq f_1 \leq f_{1+}}
 \left\{x_1 f_1 +\Sigma_{1}(f_1|x_2) \right\}.
 \label{g-x1x2}
\ee  
 The bounds of values of $f_1$ are determined by the constraint 
 $\Sigma_{1}(f_1|x_2)\geq 0$. 
 Hence, we can calculate the first-hierarchy complexity
 $\Sigma_1(f_1|x_2)$ by 
\be 
 \Sigma_{1}(f_1|x_2)=\min_{x_1}\left\{ g_1(x_1,x_2)-x_1 f_1 \right\}.
 \label{Sigma1}
\ee
 Using $\Sigma_1(f_1|x_2)$, 
 we can construct the generating function of the original complexity 
 $\Sigma(f)$, $g_2(x_2)$, which is obtained by 
\be 
 g_2(x_2) = \max_{f_{1-}\leq f_1 \leq f_{1+}}
 \left\{ f_1 +\Sigma_1(f_1|x_2) \right\}.
 \label{g_2-x}
\ee 
 There are two noteworthy points concerning this equation. 
 First, the expression of $g_2(x_2)$ is apparently the same as $g_1(1,x_2)$. 
 However, there is one difference we should notice in the evaluation
 of first-hierarchy complexity $\Sigma_1(f_1|x_2)$:
 we should carefully deal with $g_1(x_1,x_2)$ 
 derived from the 2RSB solution (see (\ref{f_2RSB-g})) 
 since it includes inappropriate branches 
 which lead to negative $\Sigma_1(f_1|x_2)$. 
 The basic line of the above procedures 
 using (\ref{Sigma1}) and (\ref{g_2-x}) is 
 correct even with such a difference.
 This point will be clearer when we apply the complexity analysis
 to the GREM in section~\ref{apply}.
 The second point is the reason why $g_2(x_2)$ becomes the generating 
 function of complexity. 
 This can be understood by putting $x_1=1$ in (\ref{g_1}). 
 In the case of $x_1=1$, the discrimination of hierarchies 1 and 2 vanishes
 and the summation runs over all pure states equally, 
 which elucidates the fact that $g_2(x_2)$ is indeed 
 the generating function of complexity.

 Once we get $g_2(x_2)$, complexity $\Sigma(f)$ can be 
 assessed in the same way as the 1RSB case: 
\be
 \Sigma(f)=\min_{x_2}\left\{ \beta x_2 f+g_2(x_2)\right\}.
 \label{Sigmaf}
\ee
 This completes procedures for evaluating complexity $\Sigma(f)$. 
 The equilibrium free energy is again evaluated from 
 $\Sigma(f)$ by (\ref{f_eq}).
 
 We here summarize the procedures to obtain complexity in the 
 2RSB level.
\begin{enumerate}
\item 
 Calculate the 2RSB solution $f_{\rm 2RSB}(m_1,m_2)$
 and introduce the generating function $g_1(x_1,x_2)$ 
 by (\ref{f_2RSB-g}).
\item 
 Calculate the first-hierarchy complexity $\Sigma_{1}(f_1|x_2)$ 
 by (\ref{Sigma1}).
\item 
 Calculate the generating function of complexity $g_{2}(x_2)$
 by (\ref{g_2-x}).
\item 
 Calculate complexity $\Sigma(f)$ by (\ref{Sigmaf}).
\end{enumerate}

 Note that quantities depending on states, 
 such as $f_1(x_2|\gamma_1)$ and $\widehat{\Sigma}_2(x_2|\gamma_1)$,
 do not appear explicitly in these procedures. It is natural 
 since those quantities cannot be calculated from the averaged 
 quantities.  

 The procedures investigated in this subsection 
 can be generalized to the $k$-step RSB with arbitrary $k$. 
 In such a case, 
 we should introduce the $i$th hierarchy complexity $\Sigma_{i}$ 
 for $1 \leq i\leq k $.
 We start from the $k$-step RSB solution 
 being the first generating function $g_1$. 
 The $i$th complexity $\Sigma_{i}$ is derived from $g_i$ as 
 (\ref{Sigma1}) and the $(i+1)$st generating function $g_{i+1}$ 
 is calculated from $\Sigma_i$ as (\ref{g_2-x}). 
 These procedures are continued from $i=1$ to $k$, and 
 the $k$th complexity corresponds to the original complexity $\Sigma(f)$.
 The Parisi's breaking parameters $\{m_i\}$ are generally related to 
 the parameters controlling the $i$th complexity, $\{x_i\}$, 
 as $x_i=m_i/m_{i+1}$ with $x_k=m_k$.

%%%%%%%%%%%%%%%%%%%%%%%%%%%%%%%%%%%%%%%%%%%%%%%%%%%%%%%%%%%%%%%%%%%%%%%%%%%%%%%%%%
\subsection{Application to the GREM}
\label{apply} 

 We saw in sections~\ref{energy} and \ref{spin} that 
 the GREM with $K=2$ can be described by the 2RSB ansatz.
 Hence, the 2RSB formulation of complexity can be 
 applied straightforwardly.

 According to (\ref{phi2rsb}) and (\ref{f_2RSB-g}), the generating function 
 of the GREM is given by
\be
 g_{1}(x_1,x_2)=\ln \alpha_{1}+\frac{\beta^2J^2}{4}a_1 x_1^2 x_2^2
 +x_1 \left( \ln \alpha_2 +\frac{\beta^2J^2}{4}a_2 x_2^2 \right).
 \label{g_1-GREM}
\ee
 The first-hierarchy complexity $\Sigma_1(f_1|x_2)$ is then 
 calculated from (\ref{Sigma1}) as 
\be
 \Sigma_{1}(f_1|x_2)=\ln \alpha_1 -\frac{1}{\beta^2 J^2a_1x_2^2}
 \left\{f_1-\left(\ln\alpha_2+\frac{\beta^2J^2}{4}a_2x_2^2\right)\right\}^2.
 \label{Sig1}
\ee
 For this calculation, the following $x_1$-parameterized forms, 
 which are valid if $\Sigma_1$ is convex and analytic,
 are quite useful:
\be
 & & f_1(x_1|x_2)=\frac{\partial g_1}{\partial x_1}, \\
 & & \Sigma_{1}(x_1|x_2)=g_1(x_1,x_2)-x_1f_1(x_1|x_2).
 \label{Sig_1-prmt}
\ee
 By the constraint $\Sigma_1(f_1|x_2)\geq 0$, 
 we find the possible range of $f_1$:
\be 
 \left|f_{1} -\left(\ln\alpha_2
 +\frac{\beta^2J^2}{4}a_2x_2^2 \right) \right|
 \leq \beta J x_2 \sqrt{a_1 \ln \alpha_1}.
\ee
 Equation (\ref{g_2-x}) gives the generating function of complexity
\be 
 g_2(x_2)=\max_{f_{1-} \leq  f_1\leq f_{1+} } 
 \left\{
 f_1+\ln \alpha_1-\frac{\left\{f_1
 -\left(\ln\alpha_2+\frac{\beta^2J^2}{4}a_2x_2^2\right)\right\}^2}
 {\beta^2J^2a_1x_2^2} \right\}. 
 \label{g_2-GREM}
\ee
 The behavior of $g_2(x_2)$ changes depending on the hierarchy parameters.
 
 For the case $T_2<T_1$, we have
\be
 g_2(x_2)=\left\{\ba{ll} 
 \ln 2+\frac{\beta^2 J^2}{4}x_2^2 & 
 0\leq x_2 \leq \frac{\beta_1}{\beta} \\
 \ln \alpha_2 +\frac{\beta^2J^2}{4} a_2 x_2^2
 + \beta J x_2 \sqrt{a_1 \ln \alpha_1} &
 \frac{\beta_1}{\beta} \leq x_2
 %\leq \frac{\beta_2}{\beta}
 \ea\right..\label{g_2-1}
\ee
 Clearly, $g_2 (x_2)$ is the same as $g_1(1,x_2)$ in (\ref{g_1-GREM})
 for small $x_2$ but different for large $x_2$. 
 This is because for large $x_2$ the maximizer in (\ref{g_2-GREM})
 should be fixed to the point $f_{1+}$ where $\Sigma_1(f_{1+}|x_2)=0$,
 but such a freezing effect in hierarchy 1 is not reflected 
 in (\ref{g_1-GREM}). 
 This leads to an inappropriate branch of $g_1(1,x_2)$. 
 
\begin{figure}[htb]
\begin{center}
\includegraphics[width=0.5\columnwidth]{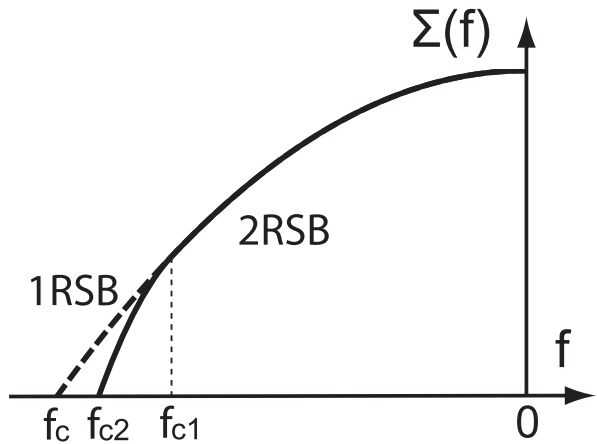}
\caption{An example of complexity behavior (\ref{Sigma-T1>T2}) for $T_2<T_1$. 
 The 1RSB solution is drawn by the dashed line.
 Both the 1RSB and the 2RSB solutions
 give the same result in the region $f>f_{c1}$.
 For the case $T_1\leq T_2$, $f_{c1}$ becomes smaller than $f_c$ and
 the 1RSB solution becomes exact in all the range of $f$.
}
\label{Sigma}
\end{center}
\end{figure}

 The complexity corresponding to (\ref{g_2-1}) 
 is assessed by (\ref{Sigma}) as
\be
 \Sigma(f)=\left\{\ba{ll}
 \ln 2 - (f/J)^2 & f_{\rm c1} \leq f \leq 0 \\ 
 \ln \alpha_2 - \frac{1}{a_2}(f/J+\sqrt{a_1 \ln \alpha_1})^2 
 & f_{\rm c2}\leq f \leq f_{\rm c1} \ea\right.,
\label{Sigma-T1>T2}
\ee
 where $f_{\rm c1}=- J\sqrt{\ln \alpha_1/ a_1}$
 and $f_{\rm c2}=- J(\sqrt{a_1\ln\alpha_1}+\sqrt{a_2\ln\alpha_2})$. 
 The lower bound $f_{\rm c2}$ yields the vanishing point of complexity 
 as $\Sigma(f_{\rm c2})=0$.
 This reproduces (\ref{f-T1>T2}) as the equilibrium free energy.

 For the case $T_1 \leq T_2$, we also have (\ref{g_2-1}) 
 from (\ref{g_2-x}), but the large-$x_2$ branch of the function 
 is inappropriate. 
 This becomes clear by calculating the resultant complexity 
\be
 \Sigma(f)=\ln 2 - \left(\frac{f}{J}\right)^2,
 \label{Sigma-T1<T2}
\ee
 for $f_{\rm c}=-J\sqrt{\ln 2} \leq f \leq 0$,
 where $f_{\rm c}$ is the vanishing point of complexity 
 $\Sigma(f_{\rm c})=0$.
 The inequality $f_{\rm c1}<f_{\rm c}$ 
 means the irrelevance of the large-$x_2$ branch of $g_{2}(x_2)$.
 This result rederives the standard REM result (\ref{f-T1<T2}), which 
 can be derived by the 1RSB ansatz, 
 as the equilibrium free energy. 
 Hence, all the results for the GREM 
 are correctly reproduced by the complexity analysis.

 Comparing the results (\ref{Sigma-T1>T2}) and (\ref{Sigma-T1<T2}), 
 we can find that the 1RSB ansatz gives the correct result for small $|f|$
 even in the case $T_2<T_1$. 
 This means that the 2RSB solution gives the same result as the 1RSB one
 as long as the first-hierarchy complexity is positive, i.e. 
 the maximizer in (\ref{g_2-GREM}) is given by a value $f_{1}<f_{1+}$. 
 We depict this in figure~\ref{Sigma}.  
%%%%%%%%%%%%%%
%%%%%%%%%%%%

%%%%%%%%%%%%%%%%%%%%%%%%%%%%%%%%%%%%%%%%%%%%%%%%%%%%%%%%%%%%%%%%%%%%%%%%%%%%%%%%%%
\subsection{Implications to Parisi's breaking parameters}

 Before closing this section, we present some arguments about the 
 breaking parameters in the Parisi solution. 
 In the 1RSB formulation, for high temperature, 
 substitution of $m=1$ in $f_{\rm 1RSB}(m)$ gives the correct equilibrium 
 free energy.
 On the other hand, at low temperature, 
 the breaking parameter $m$ is set 
 as $m=m^{*} \leq 1$, which is determined by 
 $\left. \partial f_{\rm 1RSB}(m)/\partial m\right|_{m=m^{*}}=0$.
 This condition corresponds to the vanishing point of complexity
 as written in \cite{Monasson}. 
 
 In the 2RSB formulation the same situation occurs as well.
 For the first breaking parameter $m_1$, 
 such a situation can be easily seen by using the parameterized form of 
 $\Sigma_1$ (\ref{Sig_1-prmt}), and the correspondence 
 between $f_{\rm 2RSB}(m_1,m_2)$ and $g_1(x_1,x_2)$ (\ref{f_2RSB-g}).
 In contrast, with respect to $m_2$, some delicate points emerge. 
 In terms of the generating function $g_1$, 
 the extremization condition of $f_{\rm 2RSB}(m_1,m_2)$ with 
 respect to $m_2$ gives the condition 
\be
 \left.\left( 
 \frac{\partial g_1}{\partial x_1}
 -\frac{x_2}{x_1}\frac{\partial g_1}{\partial x_2}
 \right)\right|_{x_1=m_1/m_2, x_2=m_2}=0.
\label{m_2-extr}
\ee
 Using expression (\ref{g_1}) of $g_{1}(x_1,x_2)$, 
 we can find that the left-hand side of (\ref{m_2-extr}) 
 can be written as
\be
 \frac{\partial g_1}{\partial x_1}
 -\frac{x_2}{x_1}\frac{\partial g_1}{\partial x_2}
 =\frac{\sum_{\gamma_{1}^{*}}\widehat{\Sigma}_2(f^{*}|\gamma_{1}^{*})}
 {\sum_{ \gamma_{1}^{*}} 1},\label{sig2}
\ee
 where $\gamma_{1}^{*}$ represents the valleys in hierarchy 1 contributing 
 to the saddle point, which is uniquely determined for given $x_1$ and $x_2$.
 Also, $\widehat{\Sigma}_2(f^{*}| \gamma_{1}^{*})$ denotes complexity 
 for the free energy value $f^{*}$, which 
 is the saddle point in a valley $\gamma_{1}^{*}$. 
 This equation indicates that the extremization condition 
 with respect to $m_2$ corresponds to freeze of degrees of 
 freedom in hierarchy 2. 

 These considerations give a natural interpretation 
 to the phase transitions of the case $T_2<T_1$. 
 At $T=T_1$, the degrees of freedom in hierarchy 1 
 freeze and the phase transition from RS1-RS1 to 1RSB-RS1 occurs,
 which is expressed by the 2RSB solution 
 as $f_{\rm 2RSB}(m_1=m_{1}^*,m_2=1)$ with $m_{1}^*$ being the extremizer 
 of $f_{\rm 2RSB}$ with respect to $m_1$. 
 After that, freeze of hierarchy 2 also occurs 
 at $T=T_2$ and the system goes to the 2RSB phase described 
 by $f_{\rm 2RSB}(m_1=m_{1}^*,m_2=m_{2}^*)$ 
 with the extremizer $m_{2}^*$.  

 However, for the case $T_1 \leq T_2$, the above interpretation seems
 to give an inconsistency that the quantity (\ref{sig2}) 
 becomes negative for temperature range $T_{c}< T < T_{2}$, where 
 the equilibrium free energy is given by $f_{\rm 2RSB}(m_1=1,m_2=1)$.  
 This contradiction can be understood in the 2RSB description 
 as follows.
 The condition $m_1=m_2$ leads to $x_1=1$ in (\ref{g_1}). 
 For $x_1=1$, the discrimination between 
 the hierarchies 1 and 2 vanishes. 
 In that case, 
 $\widehat{\Sigma}_2(f|\gamma_1)$ loses its meaning of complexity 
 and can be negative as long as the total complexity
 $\Sigma(f)$ is positive. 
 Hence, in such a case, we should return to the 1RSB solution 
 by putting $m_1=m_2=m$ in the 2RSB solution 
 and again take the extremization with respect to $m$.
 This consideration enables us to derive the correct equilibrium 
 free energy in this case as well. 
 The irrelevance of the negative second hierarchy entropy $s_{2}(T)$, 
 mentioned in section~\ref{sec:SP}, can be 
 understood in the same manner.

 Besides, the above discussions give an additional restriction 
 about the relation among breaking parameters $\{ m_{i} \}$. 
 Let us suppose that we treat a $k$-step RSB system and 
 have two neighboring extremizers of breaking parameters 
 $m_{i}^{*}$ and $m_{i+1}^{*}$ holding
 $m_{i}^{*} > m_{i+1}^{*}$. The conventional Parisi solution has 
 no constructive criterion for treating such a situation, though
 it requires the condition $m_{i}^{*} \leq m_{i+1}^{*}$.
 Our current consideration implies that, in such a case, 
 we should forget 
 the extremization conditions with respect to $m_i$ and $m_{i+1}$,
 and return to the $(k-1)$-step RSB by putting $m_{i}=m_{i+1}$. 
 This is a supplemental but a new criterion in the Parisi solution.

%%%%%%%%%%%%%%%%%%%%%%%%%%%%%%%%%%%%%%%%%%%%%%%%%%%%%%%%%%%%%%%%%%%%%%%%%%%%%%%%%%
%%%%%%%%%%%%%%%%%%%%%%%%%%%%%%%%%%%%%%%%%%%%%%%%%%%%%%%%%%%%%%%%%%%%%%%%%%%%%%%%%%
\section{Application of the spin representation -- generalization of the GREM}
\label{hierarchy}

 The hierarchical structure introduced in section~\ref{spin} 
 is not specific to the REM, and we can implement such 
 hierarchy to various spin models.
 In this section,
 we introduce several solvable mean-field models 
 by referring to the spin representation of the GREM.
 We discuss that such models may have various 
 hierarchical structure in phase diagram 
 and clarify what are common properties of such hierarchical models.

%%%%%%%%%%%%%%%%%%%%%%%%%%%%%%%%%%%%%%%%%%%%%%%%%%%%%%%%%%%%%%%%%%%%%%%%%%%%%%%%%%
\subsection{Pure-ferromagnetic hierarchical Ising model}

 First we implement the hierarchical structure 
 to the simplest Ising model with pure-ferromagnetic 
 $p$-body interaction.
 We study the Hamiltonian  
\be
 H = -\frac{NJ}{2}\sum_{\nu=1}^K a_\nu
 \left(\frac{1}{N_\nu}
 \sum_{i=1}^{N_\nu}\sigma_{i}^z\right)^p
 -\Gamma\sum_{i=1}^N\sigma_i^x.
\ee
 The hierarchical structure is the same as that of (\ref{grem-s}).
 Here we add the term of transverse field $\Gamma$
 to study the quantum effect.
 We introduce the magnetizations 
 $m_\nu\sim \sum_{i=1}^{N_\nu}\sigma_{i}^z/N_{\nu}$ 
 as order parameters 
 and write the free energy per $N$:
\be
 f &=& 
 \frac{J}{2}\sum_{\nu=1}^K(p-1) a_\nu m_\nu^p 
 \no\\
 & & 
 -\frac{1}{N\beta}\ln\Tr\exp\left(
 N\beta J\sum_{\nu=1}^K \frac{a_\nu}{N_\nu} 
 \tilde{m}_\nu\sum_{i=1}^{N_\nu}\sigma_i^z
 +\beta\Gamma\sum_{i=1}^N\sigma_i^x
 \right),
\ee
 where $\tilde{m}_\nu = (p/2)(m_\nu)^{p-1}$.
 The saddle-point equations at $K=2$ read 
\be
 & & m_1 = \frac{h_1+h_2}{\sqrt{(h_1+h_2)^2+\Gamma^2}}
 \tanh \left(\beta\sqrt{(h_1+h_2)^2+\Gamma^2}\right), 
 \label{sceq1} \\
 & & m_2 = \frac{N_1}{N}m_1
 +\frac{N-N_1}{N}
 \frac{h_2}{\sqrt{h_2^2+\Gamma^2}}
 \tanh\left(\beta\sqrt{h_2^2+\Gamma^2}\right),
 \label{sceq2}
\ee
 where $h_\nu = NJa_\nu\tilde{m}_\nu/N_\nu$.

 We consider the limit $p\to\infty$. From 
 the solution of the saddle-point equations, possible
 expressions of the free energy can be summarized as follows.
\begin{enumerate}
\item{Paramagnetic (P) phase: $m_1=0$, $m_2 = 0$}
\be
 f = -\frac{1}{\beta}\ln (e^{\beta\Gamma}+e^{-\beta\Gamma}).
\ee
\item{Ferromagnetic-paramagnetic (F-P) phase: 
 $m_1=1$, $m_2=N_1/N$ ($\tilde{m}_2=0$)}
\be
 f = -\frac{J}{2}a_1-\frac{N-N_1}{N}\frac{1}{\beta}
 \ln (e^{\beta\Gamma}+e^{-\beta\Gamma}).
\ee
\item{Ferromagnetic (F) phase: $m_1=1$, $m_2=1$}
\be
 f = -\frac{J}{2}.
\ee
\end{enumerate}

\begin{center}
\begin{figure}[htb]
\begin{minipage}[h]{0.5\textwidth}
\begin{center}
\includegraphics[width=0.9\columnwidth]{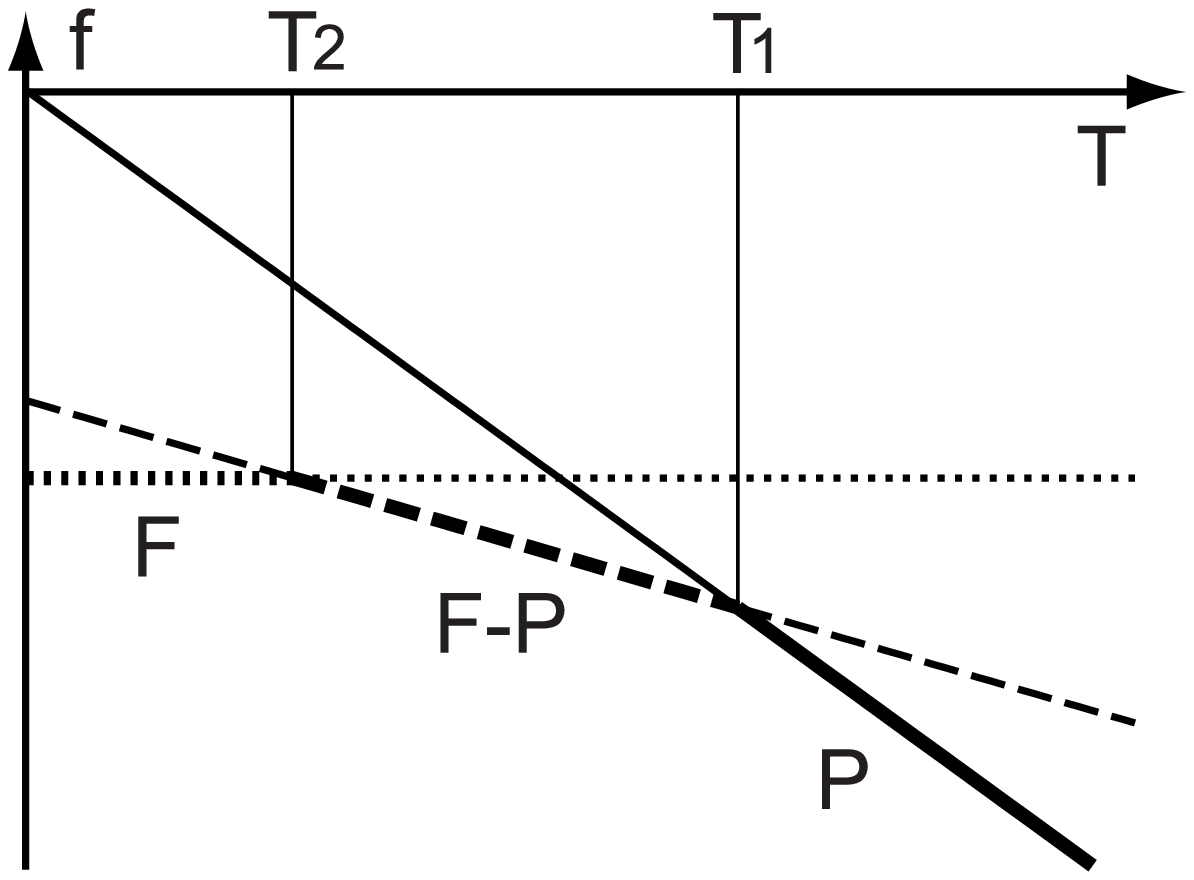}
\caption{The free energy of the 
 $K=2$ hierarchical pure ferromagnetic Ising model
 at $p=\infty$ and $\Gamma=0$ ($T_2<T_1$).
 The bold lines represent the 
 free energy chosen as appropriate solution.}
\label{fpure}
\end{center}
\end{minipage}
\begin{minipage}[h]{0.5\textwidth}
\begin{center}
\includegraphics[width=0.9\columnwidth]{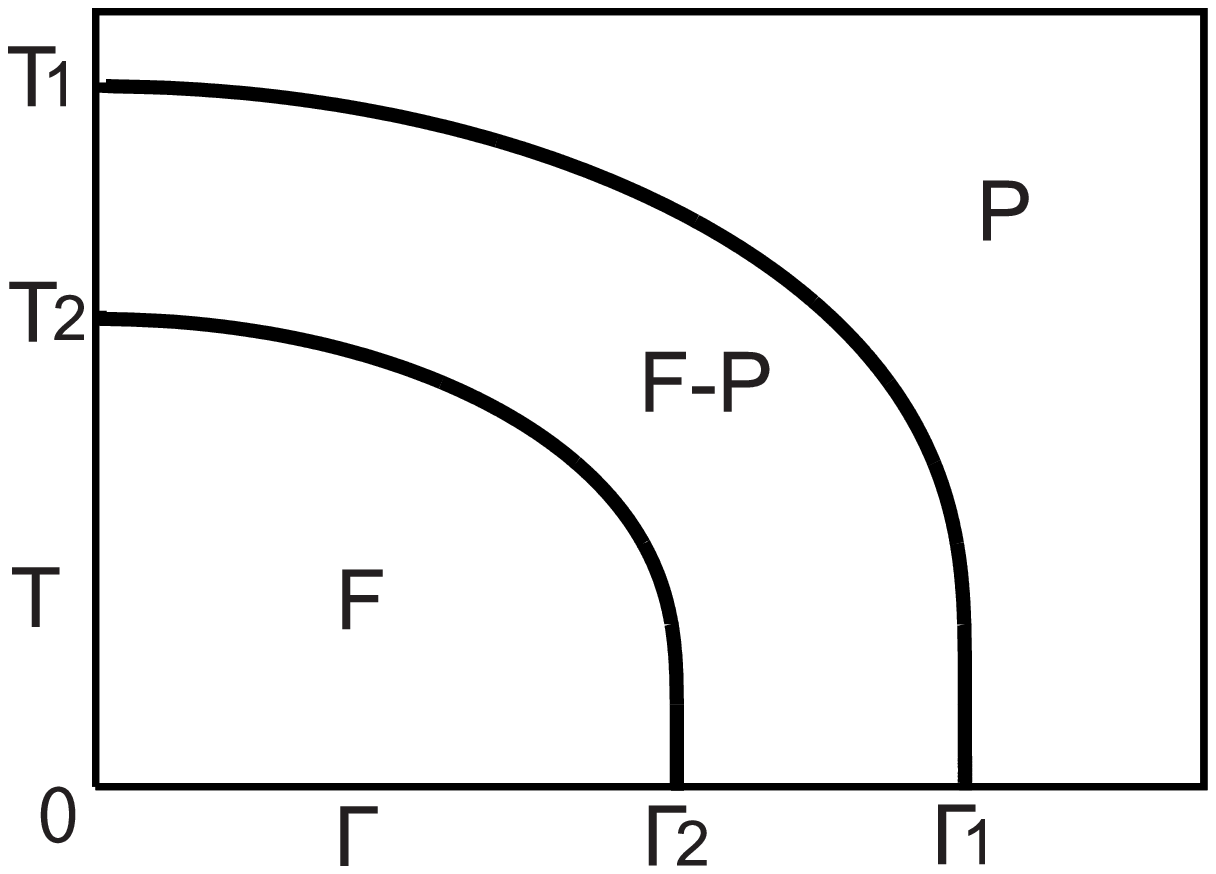}
\caption{Phase diagram of the $K=2$ hierarchical pure ferromagnetic 
 Ising model at $p=\infty$ ($T_2<T_1$).
 The critical fields at $T=0$ are given by $\Gamma_\nu = T_\nu\ln 2$.}
\label{qpure}
\end{center}
\end{minipage}
\end{figure}
\end{center}

 For $T_2<T_1$ with
\be
 T_\nu=\frac{J}{2}\frac{a_\nu}{\ln\alpha_\nu},
\ee
 the system enters the P, F-P and F phases 
 as decreasing the temperature and/or the magnitude of the transverse field.
 We depict the behavior of the free energy
 at $\Gamma=0$ in figure~\ref{fpure} 
 and the phase diagram in figure~\ref{qpure}.
 For $T_1\le T_2$ the F-P solution is irrelevant and we 
 have a single transition between P and F phases.
%%%%%%%%%%%%%%
%%%%%%%%%%%%

\begin{figure}[htb]
\begin{center}
\includegraphics[width=0.5\columnwidth]{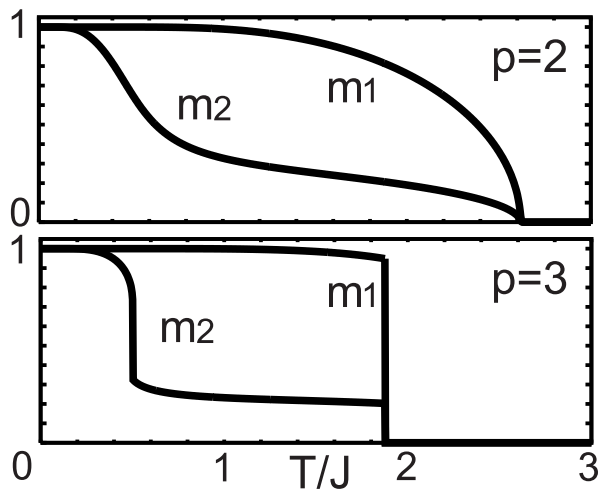}
\caption{The magnetizations $m_\nu$ for 
 the $K=2$ hierarchical Ising spin model 
 at $p=2$ and 3.
 We set parameters as $\Gamma=0$, $a_1=0.5$ and $N_1/N=0.2$.}
\label{mpure}
\end{center}
\end{figure}

 Hence a partially ordered state 
 at intermediate temperature region (F-P phase)
 is allowed in hierarchical models, which is significant
 because we have a hierarchical phase diagram even without randomness.
 We can also study the finite-$p$ case by solving 
 (\ref{sceq1}) and (\ref{sceq2}) numerically. 
 Typical behavior of magnetization in each hierarchy 
 is shown in figure~\ref{mpure}.
 In the case of $p=2$, 
 we see that there is no sharp boundary between F and F-P phases, and
 at the paramagnetic phase transition point both the magnetizations 
 go to zero at the same time.
 This behavior can generally be observed in the saddle-point equations:
 if we set $m_2=0$ in (\ref{sceq1}) and (\ref{sceq2}) we obtain $m_1=0$. 
 When $p\ge 3$, we have a discontinuous behavior
 of magnetization in each hierarchy at phase transition points.
 For instance, $m_2$ jumps from a finite value to a different finite one
 at the transition point between F and F-P phases.

%%%%%%%%%%%%%%
%%%%%%%%%%%%

%%%%%%%%%%%%%%%%%%%%%%%%%%%%%%%%%%%%%%%%%%%%%%%%%%%%%%%%%%%%%%%%%%%%%%%%%%%%%%%%%%
\subsection{GREM with ferromagnetic interaction and transverse field}

 Next we go back to the spin representation of the GREM.
 We here study the effects of the ferromagnetic interactions 
 and the transverse field on the model in section~\ref{spin}.
 The Hamiltonian is given by 
\be
 H =  -\sum_{\nu=1}^K\sum_{(i_1\cdots i_p)}^{N_\nu}
 J_{i_1\cdots i_p}^{(\nu)}\sigma_{i_1}^z\cdots\sigma_{i_p}^z
 -\frac{N}{2}\sum_{\nu=1}^KJ_{0}^{(\nu)}
 \left(\frac{1}{N_\nu}\sum_{i=1}^{N_\nu}
 \sigma_{i}^z\right)^p
 -\Gamma\sum_{i=1}^N\sigma_{i}^x. \no\\
 \label{grem-s-2}
\ee
 We consider the case where 
 the ferromagnetic interaction of each hierarchy is controlled 
 by a single parameter $J_0$ as 
\be
 J_0^{(\nu)} = J_0a_\nu.
\ee
 One of the aims to study this hierarchical ferromagnetic
 interaction is to see the relation between 
 the multi-critical points and the Nishimori line
 given for the standard REM by 
 $J_0=\beta J^2$~\cite{Nishimori, Nishimori2}.
 We are interested in how the multi-critical points 
 on the Nishimori line are located in the present hierarchical model.
 
\begin{center}
\begin{figure}[htb]
\begin{minipage}[h]{0.5\textwidth}
\begin{center}
\includegraphics[width=0.9\columnwidth]{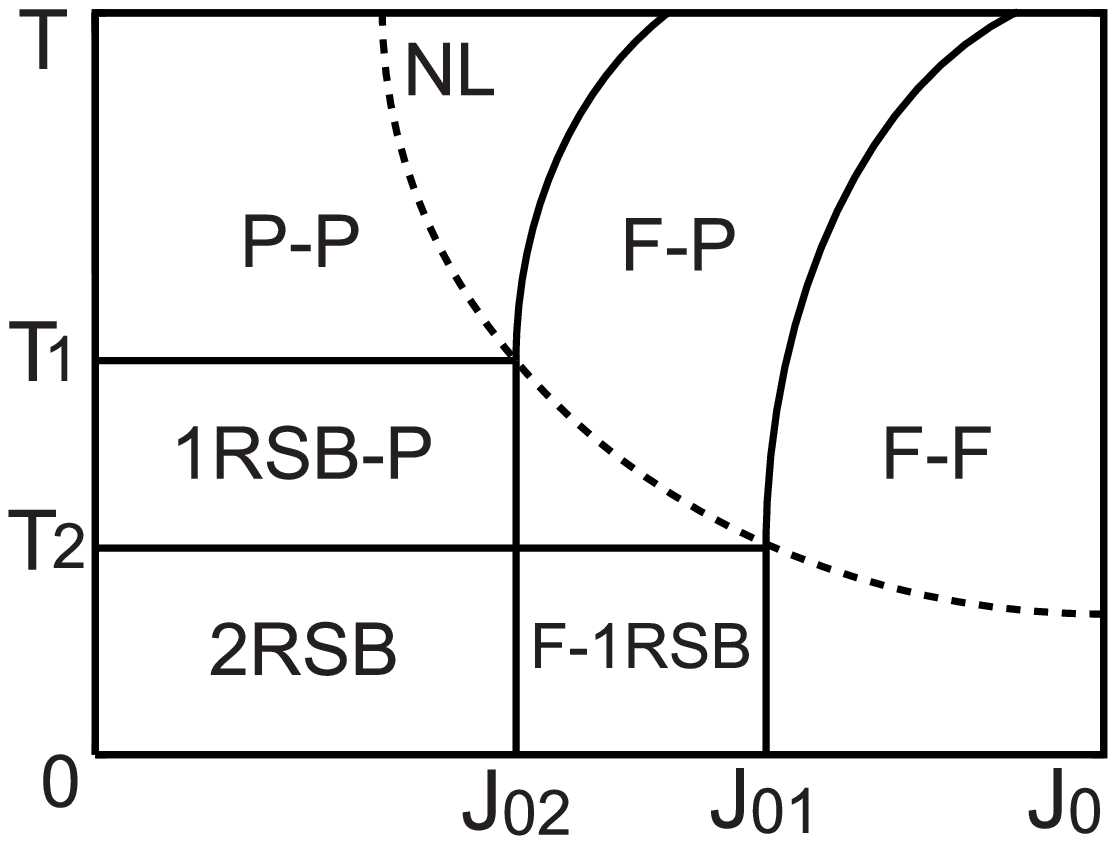}
\caption{Phase diagram of the $K=2$ GREM with ferromagnetic interaction
 ($T_2<T_1$). 
 The dashed line represents the Nishimori line $\beta J^2=J_0$.
 The critical ferromagnetic interactions at $T=0$ are given by 
 $J_{0\nu} = J^2/T_{\nu}$.}
\label{fgrem}
\end{center}
\end{minipage}
\begin{minipage}[h]{0.5\textwidth}
\begin{center}
\includegraphics[width=0.9\columnwidth]{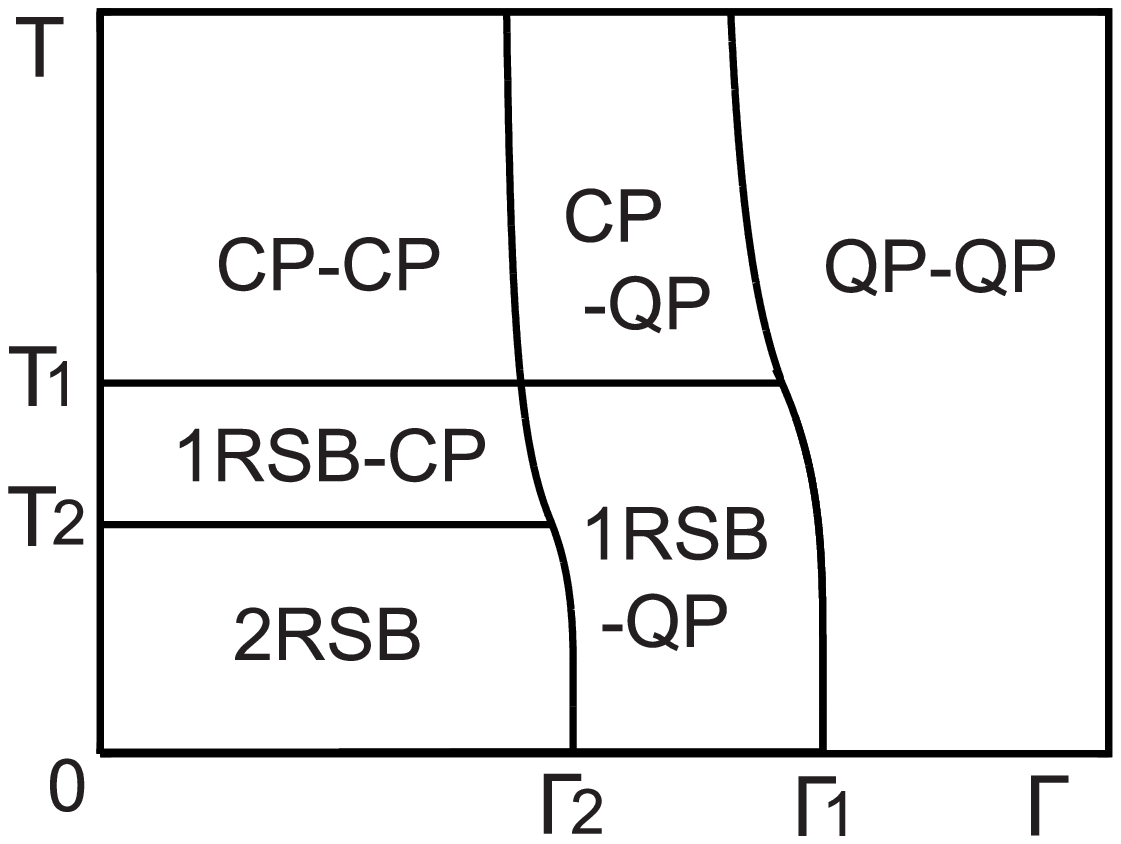}
\caption{Phase diagram of the $K=2$ GREM with transverse field
 ($T_2<T_1$).
 The critical fields at $T=0$ are given by 
 $\Gamma_{\nu} = JT_\nu/T_{\rm c}$.}
\label{tgrem}
\end{center}
\end{minipage}
\end{figure}
\end{center}

 The models without the hierarchical structure were solved 
 exactly~\cite{Goldschmidt, ONS} and it is a straightforward task 
 to generalize the method of calculation to the present case.
 The detail of the calculation is given in \ref{gremj0g}, 
 and the resultant phase diagrams 
 for $K=2$ and $T_2<T_1$ are shown in 
 figures~\ref{fgrem} (at $\Gamma=0$) and 
 \ref{tgrem} (at $J_0=0$).
 Two hierarchical structures are observed in each diagram:
 in figure~\ref{fgrem}
 multiple-step RSB and partial ferromagnetic order,
 and in figure~\ref{tgrem}
 multiple-step RSB and partial quantum effect.
 The fact that two multi-critical points are on the Nishimori line,
 as seen in figure~\ref{fgrem}, is one of the interesting results.
 In figure~\ref{tgrem}, 
 we observe a new hierarchy in two paramagnetic phases. 
 In the spin-glass model with transverse field, it is known that
 there exist two paramagnetic phases, called classical paramagnetic
 phase (denoted by CP in the figure) and quantum paramagnetic 
 phase (by QP) caused by the quantum effect of the transverse field.
 By incorporating the hierarchical structure in interactions,
 we find these two paramagnetic phases form hierarchy
 (CP-CP, CP-QP, and QP-QP phases), or there appears a 
 partially classical and partially quantum paramagnetic phase (CP-QP phase),
 as shown in the phase diagram.

%%%%%%%%%%%%%%
%%%%%%%%%%%%

%%%%%%%%%%%%%%%%%%%%%%%%%%%%%%%%%%%%%%%%%%%%%%%%%%%%%%%%%%%%%%%%%%%%%%%%%%%%%%%%%%
%%%%%%%%%%%%%%%%%%%%%%%%%%%%%%%%%%%%%%%%%%%%%%%%%%%%%%%%%%%%%%%%%%%%%%%%%%%%%%%%%%
\section{Conclusions}
\label{conclusion}

 In this paper we studied the GREM by the canonical ensemble
 formalism in conjunction with the replica method. 
 To investigate hierarchical valley structure 
 of free energy quantitatively, we generalized the notion of complexity and 
 applied it to the GREM. 
 The result not only reproduced the exact solution 
 derived by the microcanonical ensemble formalism 
 but also revealed how the higher step RSB is realized in this model,
 which is the main result of this work.

 We expect applications of the analysis of generalized complexity
 to various problems.
 For example, the replica method still has some mysteries
 in the theory itself.
 In the standard description, the full-step RSB phase,
 which corresponds to the infinite-step RSB phase, 
 is detected by a local instability of the saddle point, i.e. 
 the so-called de Almeida-Thouless condition~\cite{AT}. 
 On the other hand, according to our current scheme,
 equilibrium transitions to spin-glass phases 
 in a general $k$-step RSB system can be understood 
 as entropy crises occurring in different hierarchies. 
 The relation between these two descriptions is quite unclear. 
 We hope that the generalized complexity developed in this paper 
 will be useful to clarify this point and consequently 
 become a convenient tool in the spin-glass theory.  
 In addition, regarding multi-valley landscape of free energy,
 the method to extract information of a respective pure 
 state on adiabatic evolution
 was proposed in a recent work~\cite{ZK}. 
 In combination with their method, 
 the application of our method to
 general spin-glass models with hierarchical valley structure
 may reveal complex equilibrium/dynamical properties of spin glasses.

 We also proposed a $p$-body interacting spin-glass model 
 having a hierarchical structure in the spin sites, 
 which emulates the hierarchical structure of the GREM.
 By using the replica method, it was shown that this spin model 
 has the same thermodynamic behavior as the GREM 
 in a particular limit $p\to \infty$.
 This fact suggests that the proposed spin model exhibits 
 the higher step RSB behavior controlled by hierarchy parameters,
 though it is still analytically tractable. 
 For example, it would be interesting in the present model to examine
 the proof on the Guerra's bound of the free energy~\cite{Guerra}
 and to compare our analysis with mathematical studies \cite{BB}.
 Besides, the spin representation allows us to employ various
 methods developed for mean-field spin systems.
 One of possible applications is to incorporate
 the hierarchical structure to the TAP formulation.
 Such studies will be of great help to understand
 the higher step RSB systems.

 Another benefit of the spin representation is that 
 we can examine various physical effects on the hierarchical models. 
 As examples, we introduced ferromagnetic bias and 
 quantum effect by the transverse field to the models, 
 and analyzed both in the pure and random cases.
 The resultant phase diagrams have very rich structures, and 
 we found that partial order occurring in a part of hierarchies 
 is not specific to spin-glass phases.  
 Even in ferromagnetic phases,
 thermal and quantum fluctuations destroy the order and 
 lead to multiple-step phase transitions in different 
 hierarchies, which yields a number of multi-critical points 
 in the random systems. 
 All the multi-critical points are located on the Nishimori line, 
 which implies a universal aspect of the Nishimori line.

 As an additional application of the current work,
 the phase diagram obtained here is also useful for 
 discussion of signal processing in information theory.
 In \cite{Merhav} the relation between the GREM
 and the hierarchical random code ensemble was pointed out.
 We have applied our result to the problem of the hierarchical random
 code ensemble and shown that 
 transitions of distortion in lossy data compression
 and that of error probability in channel coding
 discussed there can be reinterpreted by the higher step RSB.
 The details will be reported soon.

%%%%%%%%%%%%%%%%%%%%%%%%%%%%%%%%%%%%%%%%%%%%%%%%%%%%%%%%%%%%%%%%%%%%%%%%%%%%%%%%%%
%%%%%%%%%%%%%%%%%%%%%%%%%%%%%%%%%%%%%%%%%%%%%%%%%%%%%%%%%%%%%%%%%%%%%%%%%%%%%%%%%%
\section*{Acknowledgments}

 The authors are grateful to Y Kabashima, T Nakajima and T Ohkubo 
 for useful discussions. 
 TO is supported by a Grant-in-Aid
 Scientific Research on Priority Areas `Novel
 State of Matter Induced by Frustration' (19052006 and 19052008).
 K Takeda is supported by
 a grant-in-aid Scientific Research on Priority Areas
 `Deepening and Expansion of Statistical Mechanical Informatics
 (DEX-SMI)' from MEXT, Japan no 18079006. 

\appendix
%%%%%%%%%%%%%%%%%%%%%%%%%%%%%%%%%%%%%%%%%%%%%%%%%%%%%%%%%%%%%%%%%%%%%%%%%%%%%%%%%%
%%%%%%%%%%%%%%%%%%%%%%%%%%%%%%%%%%%%%%%%%%%%%%%%%%%%%%%%%%%%%%%%%%%%%%%%%%%%%%%%%%

\section{RSB phases in the spin representation of the GREM}
\label{replica}

 We study the hierarchical spin model (\ref{grem-s})
 at $K=2$ and $p\to\infty$.
 We substitute possible saddle-point solutions to (\ref{zn}) and 
 show that the present model is equivalent to the GREM.
 The equivalence of the simplest RS1-RS1 solution
 has been discussed in the main body of the present paper.
 Here we study the 1RSB-RS1 and the 2RSB solutions as nontrivial cases.
 We assume $T_2<T_1$ where the critical temperatures 
 are defined in (\ref{Tnu}), 
 which is the situation
 where the phase transitions occur at $T_1$ and $T_2$.

%%%%%%%%%%%%%%%%%%%%%%%%%%%%%%%%%%%%%%%%%%%%%%%%%%%%%%
\subsection{1RSB-RS1}

 From 1RSB-RS1 solution $q_1^{ab}=\delta_m(a,b)$ and 
 $q_2^{ab}=(N_1/N)q_1^{ab}$, 
 following the standard Parisi algebra,
 we have  
 $\sum_{a>b}^n(q_1^{ab})^p=n(m-1)/2$ and 
 the second line of (\ref{zn}) as 
\be
 & & N_1\ln\Tr\exp\left(
 N\beta^2J^2\sum_{\nu=1}^2\frac{a_\nu}{N_\nu}\sum_{a>b}^n
 \tilde{q}_\nu^{ab} \sigma^a\sigma^b
 \right) \no\\
 &=& N_1\ln\Tr\exp\left\{
 \frac{\beta^2J^2}{4}\frac{N}{N_1}pa_1
 \sum_{{\rm B}}^{n/m}\left(\sum_{a\in {\rm B}}^m\sigma^a\right)^2
 -\frac{\beta^2J^2}{4}\frac{N}{N_1}pa_1 n
 \right\} \no\\
 &=&  -Nn\frac{\beta^2J^2}{4}pa_1
 +N_1\ln
 \left\{
 2\exp\left(
 \frac{\beta^2J^2}{4}\frac{N}{N_1}pa_1 m^2
 \right)
 \right\}^{n/m}.
\ee
 In the last equation 
 we take the limit $p\to\infty$
 where all spins in block ${\rm B}$ take same values $1$ or $-1$.
 Then, we obtain the result of the GREM (\ref{phinm}).

%%%%%%%%%%%%%%%%%%%%%%%%%%%%%%%%%%%%%%%%%%%%%%%%%%%%%%
\subsection{2RSB}

 Now we consider the 2RSB solution 
 $q_1^{ab}=\delta_{m_1}(a,b)$ and $q_2^{ab}=\delta_{m_2}(a,b)$.
 The 1RSB-1RSB solution can be obtained by setting $m_1=m_2$.
 The first and the third lines of (\ref{zn}) are calculated 
 in the same way as the previous case as 
\be
 & & -\frac{N\beta^2J^2}{2}(p-1)\sum_{\nu=1}^2 a_\nu\sum_{a>b}(q_\nu^{ab})^p
 +\frac{Nn\beta^2J^2}{4} \no\\
 &=& \frac{Nn\beta^2J^2}{4}\left\{
 (m_1-p(m_1-1))a_1+(m_2-p(m_2-1))a_2
 \right\}, \\
 & & (N-N_1)\ln\left\{\Tr\exp\left(
 N\beta^2J^2\frac{a_2}{N}\sum_{a>b}
 \tilde{q}^{ab}_{2} \sigma^a\sigma^b
 \right)\right\} \no\\
 &=&
 (N-N_1)n\frac{\beta^2J^2}{4}p(m_2-1)a_2+Nn\frac{\ln \alpha_2}{m_2},
\ee
 where we use the relation $(N-N_1)\ln 2=N\ln\alpha_2$
 in the second equation.
 The second line in (\ref{zn}) is written as 
\be
 & & N_1\ln\left\{\Tr\exp\left(
 N\beta^2J^2\sum_{\nu=1}^2\frac{a_\nu}{N_\nu}\sum_{a>b}^n
 \tilde{q}_\nu^{ab} \sigma^a\sigma^b
 \right)\right\} \no\\
 &=& -Nn\frac{\beta^2J^2}{4}pa_1-N_1n\frac{\beta^2J^2}{4}pa_2 
 \no\\
 & & 
 +N_1\ln\Tr\exp\left\{
 N\frac{\beta^2J^2}{4}p\sum_{\nu=1}^2\frac{a_\nu}{N_\nu}
 \sum_{{\rm B}_\nu}^{n/m_\nu}
 \left(\sum_{a\in {\rm B}_\nu}^{m_\nu}\sigma_a\right)^2
 \right\}. 
\ee
 The trace over spin variables in the last term 
 is turned out to be a formidable task
 for arbitrary $m_1$ and $m_2$. 
 Here we impose the Parisi ansatz 
 $m_1\ge m_2$ and the condition of integer $m_1/m_2$.
 Then, we can perform the trace at $p\to \infty$ as 
\be
 & & 
 N_1\ln\Tr\exp\left\{
 N\frac{\beta^2J^2}{4}p\sum_{\nu=1}^2\frac{a_\nu}{N_\nu}
 \sum_{{\rm B}_\nu}^{n/m_\nu}
 \left(\sum_{a\in {\rm B}_\nu}^{m_\nu}\sigma_a\right)^2
 \right\} \no\\
 &=& N_1\ln \left\{
 2^{n/m_1}\exp
 \left(
 N\frac{\beta^2J^2}{4}p\frac{a_1}{N_1}m_1n
 +N\frac{\beta^2J^2}{4}p\frac{a_2}{N}m_2n
 \right)
 \right\}.
\ee
 Combining everything, we finally obtain the same result (\ref{phi2rsb}) 
 as the GREM.
 
%%%%%%%%%%%%%%%%%%%%%%%%%%%%%%%%%%%%%%%%%%%%%%%%%%%%%%%%%%%%%%%%%%%%%%%%%%%%%%%%%%
%%%%%%%%%%%%%%%%%%%%%%%%%%%%%%%%%%%%%%%%%%%%%%%%%%%%%%%%%%%%%%%%%%%%%%%%%%%%%%%%%%
\section{GREM with ferromagnetic interaction and transverse field}
\label{gremj0g}

 We study (\ref{grem-s-2}).
 Under the saddle point evaluation,
 the average of the replicated partition function is written as 
\be
 [Z^n] &=&
 \exp\left\{
 -\frac{N\beta J_0}{2}(p-1)\sum_{\nu=1}^K a_\nu
 \sum_{a=1}^n(m_\nu^a)^p 
 \right.\no\\
 & & 
 -\frac{N\beta^2 J^2}{4}(p-1)\sum_{\nu=1}^K a_\nu
 \left(
 \sum_{a=1}^n(\chi_\nu^a)^p +\sum_{a\ne b}^n(q_\nu^{ab})^p\right)
 \no\\
 & & 
 +\ln\Tr \exp\left(
 N\beta J_0\sum_{\nu=1}^K\frac{a_\nu}{N_\nu}\sum_{a=1}^n\tilde{m}_\nu^a
 \sum_{i=1}^{N_\nu} S_{zi}^a
 \right.\no\\
 & & 
 +\frac{N\beta^2J^2}{2}\sum_{\nu=1}^K\frac{a_\nu}{N_\nu}
 \sum_{a=1}^n\tilde{\chi}_\nu^a\sum_{i=1}^{N_\nu}
 S_{zi}^aS_{zi}^a
 \no\\
 & & \left.\left.
 +N\beta^2J^2\sum_{\nu=1}^K\frac{a_\nu}{N_\nu}\sum_{a>b}^n
 \tilde{q}_\nu^{ab}\sum_{i=1}^{N_\nu} S_{zi}^a S_{zi}^b
 +\sum_{a=1}^{n}\Gamma \sum_{i=1}^NS_{xi}^a
 \right)\right\}.
\ee
 We use the imaginary time formalism, where
 the classical spin variables $\Vec{S}_{i}(\tau)$ 
 are defined for each imaginary time $\tau$ between 0 and $\beta$.
 $S_{zi}=(1/\beta)\int_0^\beta d\tau S_{zi}(\tau)$
 and the trace represents multiple integrals over spin variables
 with a proper measure~\cite{Takahashi}.
 The order parameters are introduced as 
 $m_\nu^a\sim \sum_{i=1}^{N_\nu}S_{zi}^a(\tau)/N_{\nu}$, 
 $\chi_\nu^a\sim
 \sum_{i=1}^{N_\nu}S_{zi}^a(\tau)S_{zi}^a(\tau')/N_{\nu}$, 
 and 
 $q_\nu^{ab}\sim 
 \sum_{i=1}^{N_\nu}S_{zi}^a(\tau)S_{zi}^b(\tau')/N_{\nu}$.
 We also use $\tilde{m}_\nu^a=(p/2)(m_\nu^a)^{p-1}$ and so on.
 The deviation of the parameter $\chi$ from unity 
 represents the magnitude of the quantum effect.
 At the limit of $p\to\infty$, 
 the RS ansatz for $m_\nu^a$ and $\chi_\nu^a$
 and the static approximation for all parameters 
 are justified~\cite{ONS}.
 By taking possible solutions into account, 
 we can classify the phases at $K=2$ as summarized below.
 We consider two cases $\Gamma=0$ and $J_0=0$
 for simplicity.

%%%%%%%%%%%%%%%%%%%%%%%%%%%%%%%%%%%%%%%%%%%%%%%%%%%%%%%%%%%%%%%%%%%%%%%%%%%%%%%%%%
\subsection{Ferromagnetic interaction}

 In the case of the classical limit $\Gamma=0$, 
 we have six phases.
 The RS1-RS1 (paramagnetic), 
 1RSB-RS1 and 2RSB phases are the same as the previous calculation
 with $J_0=0$. 
 In addition, 
 we have three new phases including $J_0$ in their free energies:
\begin{itemize}
\item{F-P: $(m_1,m_2)=(1,N_1/N)$, 
 $(q_1^{ab},q_2^{ab})=(1,\delta_{ab})$}
\be
 f = -\frac{J_0}{2}a_1
 -\frac{1}{\beta}\ln\alpha_2-\frac{\beta^2J^2}{4}a_2.
\ee
\item{F-F: $(m_1,m_2)=(1,1)$, $(q_1^{ab},q_2^{ab})=(1,1)$}
\be
 f = -\frac{J_0}{2}.
\ee
\item{F-1RSB: $(m_1,m_2)=(1,N_1/N)$,  
 $(q_1^{ab},q_2^{ab})=(1,\delta_m(a,b))$}
\be
 f = -\frac{J_0}{2}a_1-J\sqrt{a_2\ln\alpha_2}.
\ee
\end{itemize}

 Comparing these free energies, 
 we obtain the phase diagram in figure~\ref{fgrem}.

%%%%%%%%%%%%%%%%%%%%%%%%%%%%%%%%%%%%%%%%%%%%%%%%%%%%%%%%%%%%%%%%%%%%%%%%%%%%%%%%%%
\subsection{Transverse field}

 When $J_0=0$, we have phase transitions to the quantum paramagnetic
 phase with $\chi\ne 1$.
 We obtain the following three new quantum phases
 and the phase diagram in figure~\ref{tgrem}.
\begin{itemize}
\item{Classical P- Quantum P (CP-QP): 
 $(\chi_1,\chi_2)=(1,N_1/N)$,  
 $(q_1^{ab},q_2^{ab})=(\delta_{ab},\delta_{ab})$}
\be
 f = -\frac{J_0}{2}a_1
 -\frac{1}{\beta}\ln\alpha_2-\frac{\beta^2J^2}{4}a_2.
\ee
\item{QP-QP: 
 $(\chi_1,\chi_2)=(\tanh\beta\Gamma/\beta\Gamma,\tanh\beta\Gamma/\beta\Gamma)$,  
 $(q_1^{ab},q_2^{ab})=(\delta_{ab},\delta_{ab})$}
\be
 f = -\frac{1}{\beta}\ln\left(e^{\beta\Gamma}+e^{-\beta\Gamma}\right).
\ee
\item{1RSB-QP: $(\chi_1,\chi_2)=(1,N_1/N)$,  
 $(q_1^{ab},q_2^{ab})=(\delta_m(a,b),\delta_{ab})$}
\be
 f = -J\sqrt{a_1\ln\alpha_1}
 -\frac{N-N_1}{N}\frac{1}{\beta}\ln\left(e^{\beta\Gamma}+e^{-\beta\Gamma}\right).
\ee
\end{itemize}

%%%%%%%%%%%%%%%%%%%%%%%%%%%%%%%%%%%%%%%%%%%%%%%%%%%%%%%%%%%%%%%%%%%%%%%%%%%%%%%%%%
%%%%%%%%%%%%%%%%%%%%%%%%%%%%%%%%%%%%%%%%%%%%%%%%%%%%%%%%%%%%%%%%%%%%%%%%%%%%%%%%%%

\end{document}